\documentclass[aps,prc,superscriptaddress,twocolumn,nofootinbib,floatfix,amssymb,preprintnumbers,amsmath]{revtex4-1}

\usepackage{graphicx}
\usepackage{dcolumn}
\usepackage{bm}
\newcommand{\braket}[1]{\left\langle #1 \right\rangle}
\newcommand{\bra}[1]{\left \langle #1 \right|}
\newcommand{\ket}[1]{\left| #1 \right \rangle}
\usepackage{bm, color}
\usepackage{lipsum}
\usepackage{float}

\definecolor{ryangreen}{rgb}{0.81,0.13,0.16}

\definecolor{red}{rgb}{0.81,0.13,0.16}

\usepackage{hyperref}
\hypersetup{
    colorlinks=true,
    linkcolor=blue,
    filecolor=magenta,      
    urlcolor=blue,
    citecolor=blue,
    }

\begin{document}

\preprint{LA-UR-24-30073}

\title{Perturbative treatment of nonlocal chiral interactions in auxiliary-field diffusion Monte Carlo calculations}

\author{Ryan Curry}
\affiliation{Department of Physics, University of Guelph, Guelph, Ontario N1G 2W1, Canada}
\affiliation{Theoretical Division, Los Alamos National Laboratory, Los Alamos, New Mexico 87545, USA}
\author{Rahul Somasundaram}
\affiliation{Theoretical Division, Los Alamos National Laboratory, Los Alamos, New Mexico 87545, USA}
\affiliation{Department of Physics, Syracuse University, Syracuse, NY 13244, USA}
\author{Stefano Gandolfi}
\affiliation{Theoretical Division, Los Alamos National Laboratory, Los Alamos, New Mexico 87545, USA}
\author{Alexandros Gezerlis}
\affiliation{Department of Physics, University of Guelph, Guelph, Ontario N1G 2W1, Canada}
\author{Ingo Tews}
\affiliation{Theoretical Division, Los Alamos National Laboratory, Los Alamos, New Mexico 87545, USA}

\date{\today}

\begin{abstract}
Nuclear many-body systems, ranging from nuclei to neutron stars, are some of the most interesting physical phenomena in our universe, and Quantum Monte Carlo (QMC) approaches are among the most accurate many-body methods currently available to study them. 
In recent decades, interactions derived from chiral effective field theory (EFT) have been widely adopted in the study of nuclear many-body systems. 
One drawback of the QMC approach is the requirement that the nuclear interactions need to be local, whereas chiral EFT interactions usually contain nonlocalities.
In this work, we leverage the capability of computing second-order perturbative corrections to the ground-state energy in order to develop a self-consistent approach to including nonlocal operators in QMC calculations.
We investigate both the deuteron and the neutron-matter equation of state in order to show the robustness of our technique, and pave the way for future QMC calculations at higher orders in the EFT, where nonlocal operators cannot be avoided. 
\end{abstract}

\maketitle


\section{Introduction}

Understanding the forces between nucleons is one of the key pieces necessary to study the nuclear many-body problem \cite{Bethe_1971}. 
At a fundamental level,  nucleon-nucleon interactions are fully described by the theory of Quantum Chromodynamics (QCD) in the language of quarks and gluons. 
However, QCD is not tractable in the low-energy regime of nuclear physics \cite{Drischler_Haxton_McElvain_etal_2021}. 
To circumvent this issue, modern nuclear interactions are derived from chiral effective field theory (EFT) following the work by Weinberg and others \cite{Weinberg_1979, Weinberg_1990, Weinberg_1991, Ordonez_Ray_vanKolck_1994, Ordonez_Ray_vanKolck_1996, vanKolck_1994, Georgi_1993}.
Chiral EFT allows nuclear many-body calculations to be concerned only with the relevant degrees of freedom, i.e., nucleons and pions. 
In this framework, nuclear interactions are written down in terms of short-range contact operators as well as intermediate- and long-range pion exchanges \cite{Epelbaum_Hammer_Meissner_2009, Machleidt_Entem_2011}. 

Nuclear interactions derived from chiral EFT have been extraordinarily successful in a wide range of configuration-space nuclear many-body methods, investigating nuclei from many regions of the nuclear chart \cite{Hagen_Papenbrock_Dean_etal_2008, Sun_Morris_Hagen_etal_2018, Barrett_Navratil_Vary_2013, Stroberg_Holt_Schwenk_etal_2021}. 
The issue is complicated however, if one is interested in using a quasi-exact many-body method like the quantum Monte Carlo (QMC) family of approaches~\cite{Carlson_1987, Schmidt_Fantoni_1999, Carlson_Gandolfi_Pederiva_etal_2015}. 
Nucleon-nucleon interactions are considered local if they only depend on the radial separation between particles, which can be translated into a dependence on the momentum transfer between the nucleons.  
However, due to their natural casting as a momentum space expansion, chiral nucleon-nucleon interactions typically depend on this additional momentum scale, and hence, are nonlocal~\cite{Epelbaum_Glockle_Meissner_2005, Entem_Machleidt_2003}.
They employ both nonlocal regulators as well as nonlocal contact operators. 
This posed a significant challenge for implementation in a QMC approach, since almost all QMC methods are cast in coordinate space and require local interactions as input. 
They struggle to handle nonlocal operators in the imaginary time propagator, due to the stochastic nature of the calculation, i.e., sampling from the potential propagator with derivative terms quickly drowns the calculation in noise, and the propagator is no longer guaranteed to be positive definite~\cite{Lynn_Schmidt_2012,Carlson_Gandolfi_Pederiva_etal_2015}.   
Several attempts to deal with the nonlocal chiral EFT operators involved the development of nonlocal propagators \cite{Lynn_Schmidt_2012} or of novel QMC algorithms which avoided coordinate space \cite{Roggero_Mukherjee_Pederiva_2013}. 
More successfully, local versions of the chiral EFT interactions up to next-to-next-to-leading order (N$^2$LO) were constructed \cite{Gezerlis_Tews_Epelbaum_etal_2013, Gezerlis_Tews_Epelbaum_etal_2014, Tews_Gandolfi_Gezerlis_etal_2016}.
These interactions take advantage of the fact that the long- and intermediate-range pion exchanges can be written in a completely local form up to N$^2$LO, depending on the power counting used \cite{Weinberg_1991}, and that the short-range contact operators can also be chosen to be local as a result of the required antisymmetry of fermionic wavefunctions \cite{Gezerlis_Tews_Epelbaum_etal_2013}. 
Maximally local versions of the chiral EFT interactions have also been developed up to next-to-next-to-next-to-leading order (N$^3$LO), with \cite{Piarulli_Girlanda_Schiavilla_etal_2015, Piarulli_Girlanda_Schiavilla_etal_2016} and without \cite{Saha_Entem_Machleidt_etal_2023, Somasundaram_Lynn_Huth_etal_2024} $\Delta$-isobars.
At N$^3$LO, however, it is impossible to derive completely local interactions and nonlocal operators cannot be avoided. 
Therefore, future QMC calculations employing a chiral EFT interaction at N$^3$LO or beyond must be able to handle the inclusion of nonlocal operators. 

In this work, we employ the auxiliary-field diffusion Monte Carlo (AFDMC) method~\cite{Schmidt_Fantoni_1999} and implement chiral EFT interactions containing nonlocal operators.
Our approach is to treat local parts of the interaction in the propagator (see \ref{QMC} below) and to include any nonlocal terms perturbatively.
We limit ourselves to using chiral EFT interactions at next-to-leading order (NLO) to avoid treating many-nucleon forces that enter at N$^2$LO and beyond. We include the nonlocal operators perturbatively at second-order following Ref.~\cite{Curry_Lynn_Schmidt_etal_2023}, and study both the deuteron and neutron matter ground-state energies to show the robustness of our technique. 

\section{Nuclear Hamiltonians and Chiral Effective Field Theory} \label{NuclearH}

One of the strengths of QMC methods is that a single Hamiltonian can be used to describe diverse systems such as atomic nuclei, as well as pure infinite neutron matter that is very similar to the neutron-rich matter in the 
outer cores of neutron stars. 
The nuclear many-body Hamiltonian can be written as
\begin{align}
    H = -\frac{\hbar^2}{2m} \sum_i^N \nabla_i^2 + \sum_{i<j} V(r_{ij}) + ...\,,
\end{align}
where $N$ is the total number of nucleons and the two-body potential $V(r_{ij})$ acts between all pairs of particles.
The explicit form of the Hamiltonian has been given in a variety of ways, due to the fact that nuclear-structure is agnostic to the short-range details of the nuclear interaction \cite{Bogner_Furnstahl_Schwenk_2010}.  
Modern \textit{ab initio} calculations employ nuclear Hamiltonians derived from chiral EFT \cite{Epelbaum_Hammer_Meissner_2009, Machleidt_Entem_2011}. 
In principle, the nuclear Hamiltonian includes \mbox{three-,} {four-,} and higher many-body interactions.
As we limit ourselves to a chiral EFT interaction including terms up to NLO, we are able to neglect these terms in this study.

Modern chiral EFT interactions represent a natural choice for QMC calculations, due to the fact they offer a consistent, systematically improvable framework for nuclear forces.
This is because chiral EFT describes nuclear interactions as an expansion in powers of $Q/\Lambda_b$, where $Q$ is the momentum of the nucleons and $\Lambda_b$ is the breakdown scale of the expansion. 
The interactions can then be written as a series of terms at successfully higher chiral orders,
\begin{align}
V = V^{(0)} + V^{(2)} + V^{(3)} + ... \,.
\end{align}
As mentioned above, chiral EFT interactions are often divided into long- and intermediate-range contributions due to pion exchanges and short-range contact interactions. 
Depending on the power counting scheme used, the pion-exchange terms (up to N$^2$LO) can be written entirely in terms of the momentum transfer, $\bm{q}=\bm{p}^{\prime}-\bm{p}$, where $\bm{p}$ and $\bm{p}^{\prime}$ are the incoming and outgoing relative momenta. 
This means they can be expressed in an entirely local form which is easily amenable to QMC calculations \cite{Gezerlis_Tews_Epelbaum_etal_2013}.  
The contact terms, which are the focus of this work, are slightly more involved.
Different contact operators enter chiral EFT interactions at different orders and will be discussed below.
Furthermore, when implementing EFT interactions in many-body computational algorithms, one needs to  impose a regulator function to control the interactions at high momenta, that otherwise would lead to the appearance of divergences~\cite{Machleidt_Entem_2011}. 
Local chiral EFT interactions employ local regulators that smooth out the interactions at very short distances \cite{Gezerlis_Tews_Epelbaum_etal_2013}. 
For our work, we employ the local regulators of Ref.~ \cite{Somasundaram_Lynn_Huth_etal_2024, Tews_Somasundaram_Lonardoni_etal_2024} which are parameterized by a coordinate-space cutoff $R_0$.
This scale can be viewed as a measure of how hard the core of the potential is, with smaller values of $R_0$ leading to harder interactions.

At LO, the most general set of contact terms, $\{ 1, \bm{\sigma}_1 \cdot \bm{\sigma}_2, \bm{\tau}_1 \cdot \bm{\tau}_2, \bm{\sigma}_1 \cdot \bm{\sigma}_2\bm{\tau}_1 \cdot \bm{\tau}_2 \}$ is momentum independent. 
Due to the fact we are interested in fermionic systems, which means we will be dealing with antisymmetric wavefunctions, only two of these contact operators need to be included in the interaction.
The remaining operators are recovered through antisymmetrization. 
To see this, one can antisymmetrize the interaction containing all four LO contact terms and find that only two of the terms are linearly independent \cite{Gezerlis_Tews_Epelbaum_etal_2014}. 
As a result, current state-of-the-art local chiral EFT interactions \cite{Gezerlis_Tews_Epelbaum_etal_2014, Somasundaram_Lynn_Huth_etal_2024} choose to write $V^{(0)}_{\text{cont}} = C_S + C_T \bm{\sigma}_1 \cdot \bm{\sigma}_2$ for their LO interaction. 

\begin{figure*}
\centering
\includegraphics[width=1\textwidth]{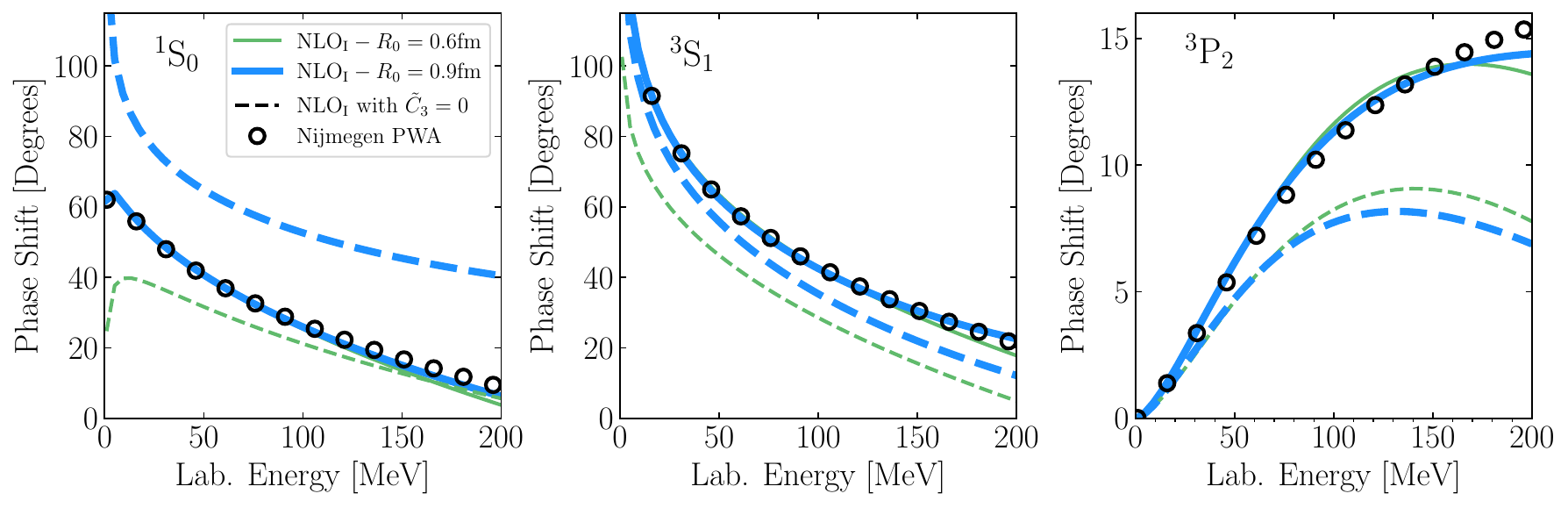} 
   \caption{The LO and NLO LECs from our NLO$_{\text{I}}$ interaction are fit to Nijmegen partial wave analysis data \cite{Stoks_Klomp_Rentmeester_etal_1993} in a number of neutron-proton scattering channels, shown here for two values of the coordinate space cutoff. The solid lines are obtained while including the nonlocal operator. However, since our many-body calculations can only treat this operator perturbatively, we recompute the phase shift values for our interaction assuming the nonlocal LEC $\tilde{C}_3=0$ (plotted with dashed lines) to give a measure of the physics missing from our non-perturbative calculations.}
\label{fig:phase_shifts}
\end{figure*}

At NLO,  employed in this work, the situation is more complicated. 
There are a total of 14 different contact operators allowed by symmetries,
\begin{align}
    V_{\text{cont}}^{(2)} = &\gamma_1 q^2 + \gamma_2 q^2 \bm{\sigma}_1\cdot \bm{\sigma}_2 + \gamma_3 q^2 \bm{\tau}_1\cdot\bm{\tau}_2 \nonumber
    \\
    &+\gamma_4q^2  \bm{\sigma}_1\cdot \bm{\sigma}_2\bm{\tau}_1\cdot\bm{\tau}_2 \nonumber
    \\
    &+ \gamma_5 k^2 + \gamma_6 k^2 \bm{\sigma}_1\cdot \bm{\sigma}_2 + \gamma_7 k^2 \bm{\tau}_1\cdot\bm{\tau}_2 \nonumber
    \\
    &+ \gamma_8 k^2 \bm{\sigma}_1\cdot \bm{\sigma}_2\bm{\tau}_1\cdot\bm{\tau}_2 + \gamma_9(\bm{\sigma}_1 + \bm{\sigma}_2)\cdot(\bm{q} \times \bm{k}) \nonumber
    \\
    &+ \gamma_{10}(\bm{\sigma}_1 + \bm{\sigma}_2)\cdot(\bm{q} \times \bm{k}) \bm{\tau}_1\cdot\bm{\tau}_2 + \gamma_{11}(\bm{\sigma}_1\cdot\bm{q})(\bm{\sigma}_2\cdot\bm{q}) \nonumber
    \\
    &+ \gamma_{12}(\bm{\sigma}_1\cdot\bm{q})(\bm{\sigma}_2\cdot\bm{q})\bm{\tau}_1\cdot\bm{\tau}_2 + \gamma_{13}(\bm{\sigma}_1\cdot\bm{k})(\bm{\sigma}_2\cdot\bm{k}) \nonumber
    \\
    &+ \gamma_{14}(\bm{\sigma}_1\cdot\bm{k})(\bm{\sigma}_2\cdot\bm{k})\bm{\tau}_1\cdot\bm{\tau}_2\,.
\end{align}
In a similar fashion to the case at LO, it can be shown that the NLO contribution can be written in terms of 7 linearly independent operators. 
Terms in the above list that depend only on the momentum transfer $\bm{q}$ will Fourier transform into local operators, while terms depending on the momentum transfer in the exchange channel $\bm{k}$ will transform into nonlocal operators. 
Initial chiral EFT interactions built in momentum space \cite{Epelbaum_Glockle_Meissner_2004, Entem_Machleidt_2003} chose the set of 7 operators that avoids isospin dependence, but there are more recent potentials with varying degrees of nonlocality. 
For QMC calculations \cite{Gezerlis_Tews_Epelbaum_etal_2014}, the NLO interaction is typically given by choosing the local and spin-orbit operators, 
\begin{align}
    V_{\text{cont}}^{(2)} = &C_1  q^2 + C_2 q^2 \bm{\tau}_1 \cdot \bm{\tau}_2 \nonumber
    \\
    &+ C_3 q^2 \bm{\sigma}_1\cdot\bm{\sigma}_2 + C_4q^2 \bm{\sigma}_1 \cdot \bm{\sigma}_2 \bm{\tau}_1 \cdot \bm{\tau}_2 \nonumber
    \\
    & + \frac{iC_5}{2}(\bm{\sigma}_1 + \bm{\sigma}_2)\cdot \bm{q} \times \bm{k} + C_6 (\bm{\sigma}_1 \cdot \bm{q})(\bm{\sigma}_1\cdot\bm{q}) \nonumber
    \\
    &+ C_7(\bm{\sigma}_1 \cdot \bm{q})(\bm{\sigma}_1\cdot\bm{q}) \bm{\tau}_1\cdot\bm{\tau}_2\,. 
\end{align}
Though there are no new contact operators introduced at N$^2$LO, starting at N$^3$LO it is no longer possible to avoid including nonlocal terms \cite{Piarulli_Girlanda_Schiavilla_etal_2015, Somasundaram_Lynn_Huth_etal_2024}.  

Due to the Fierz rearrangement freedom to choose different NLO contact operators, we could instead select a different set of contact terms and expect to reproduce the same physical results up to regulator artifacts~\cite{Huth_2018}.
For example, by replacing the $q^2 \bm{\sigma}_1 \cdot \bm{\sigma}_2$ operator with a nonlocal equivalent, $k^2 \bm{\sigma}_1 \cdot \bm{\sigma}_2$, we now have
\begin{align}  \label{newNLO}
    V_{\text{cont}}^{(2)*} = &\tilde{C_1}  q^2 + \tilde{C_2} q^2 \bm{\tau}_1 \cdot \bm{\tau}_2 \nonumber
    \\
    &+ \tilde{C_3} k^2 \bm{\sigma}_1\cdot\bm{\sigma}_2 + \tilde{C_4}q^2 \bm{\sigma}_1 \cdot \bm{\sigma}_2 \bm{\tau}_1 \cdot \bm{\tau}_2 \nonumber
    \\
    & + \frac{i\tilde{C_5}}{2}(\bm{\sigma}_1 + \bm{\sigma}_2) \cdot \bm{q} \times \bm{k} + \tilde{C_6} (\bm{\sigma}_1 \cdot \bm{q})(\bm{\sigma}_1\cdot\bm{q}) \nonumber
    \\
    &+ \tilde{C_7}(\bm{\sigma}_1 \cdot \bm{q})(\bm{\sigma}_1\cdot\bm{q}) \bm{\tau}_1\cdot\bm{\tau}_2\,, 
\end{align}
which is an equivalent form of the NLO contact potential. 

The choice of the $k^2 \bm{\sigma}_1 \cdot \bm{\sigma}_2$ operator as a test case is justified as it includes both nonlocal contributions in addition to spin-dependent operators that are characteristic of realistic nuclear forces.
In $V_{\text{cont}}^{(2)*}$ the majority of terms are local and can be handled easily in QMC, while a single operator is nonlocal and must be treated perturbatively. The Fourier transforms of the local operators are derived in \cite{Gezerlis_Tews_Epelbaum_etal_2013} and we derive the Fourier transform for the nonlocal operator in Appendix A. 
The full coordinate-space representation of $V_{\text{cont}}^{(2)*}$ is given by
\begin{align} \label{coordinatespace}
V^{(2)*}_{\text{cont}}(\bm{r}) = &-(\tilde{C_1} + \tilde{C_2} \bm{\tau}_1\cdot\bm{\tau}_2) \Delta \delta(\bm{r}) \nonumber
\\
&-\tilde{C_4}\bm{\tau}_1 \cdot \bm{\tau}_2\bm{\sigma}_1 \cdot \bm{\sigma}_2 \Delta \delta(\bm{r}) +\frac{\tilde{C_5}}{2}\frac{\partial_r \delta(\bm{r})}{r} \bm{L}\cdot\bm{S} \nonumber
\\
& + (\tilde{C_6} + \tilde{C_7}\bm{\tau}_1 \cdot \bm{\tau}_2) \nonumber
\\
&\times \biggl[(\bm{\sigma}_1 \cdot \hat{\bm{r}})(\bm{\sigma}_2 \cdot \hat{\bm{r}}) \biggl(\frac{\partial_r \delta(\bm{r})}{r} - \partial_r^2 \delta(\bm{r}) \biggl) \nonumber
\\
&\ \ \ \ \ \ - \bm{\sigma}_1 \cdot \bm{\sigma}_2 \frac{\partial_r \delta(\bm{r})}{r} \biggl] \nonumber
\\
&+ \tilde{C_3} \bm{\sigma}_1 \cdot \bm{\sigma}_2 \biggl[ -\frac{1}{4}(\Delta \delta(\bm{r}))\psi(\bm{r}) \nonumber
\\
& \ \ \ \ -\frac{1}{r} \frac{\partial \delta(\bm{r})}{\partial r}\biggl(\bm{r} \cdot \nabla \psi(\bm{r})\biggl) - \delta(\bm{r})\Delta \psi(\bm{r}) \biggl]\,,
\end{align}
where $\delta(\bm{r})$ is the local regulator function employed in Ref.~\cite{Somasundaram_Lynn_Huth_etal_2024, Tews_Somasundaram_Lonardoni_etal_2024}.

In this work, we treat the NLO interaction of Eq.~(\ref{coordinatespace}) in different ways. 
First, as a test of our approach, we employ Eq. (\ref{coordinatespace}) with the nonlocal LEC set to zero and fit the remaining local LECs to the neutron-proton phase shifts using the least-squares minimization procedure described in Ref.~\cite{Somasundaram_Lynn_Huth_etal_2024}. 
Briefly, the relevant LECs that appear at LO and NLO are fit to the $^1$S$_0$, $^3$S$_1$, $\epsilon_1$, $^1$P$_1$, $^3$P$_0$, $^3$P$_1$, and $^3$P$_2$ partial waves up to a lab energy of $150$~MeV, where the experimental data are taken to be the phase-shift values from the Nijmegen partial-wave analysis (NPWA)~\cite{Stoks_Klomp_Rentmeester_etal_1993}.
We then leave the local LECs ``frozen" and tune the nonlocal LEC over a range of values. 
This allows us to explore the perturbativeness of the nonlocal operator and test our AFDMC approach for the nonlocal contribution to our Hamiltonian. 
Since the nonlocal term is set to zero while fitting this interaction, this ``frozen" NLO interaction will be denoted by NLO$_0$.
The second, more realistic, approach is to ``refit" the interaction to the phase shifts while including the nonlocal LEC in the fit.
Interactions NLO$_{\text{I}}$ and NLO$_{\text{II}}$, discussed in more detail in Section \ref{deuteron}, employ this approach. 

By design, including the nonlocal operator in the phase-shift fit means that the imaginary-time propagator in our AFDMC calculation will be missing information about the interaction, since the nonlocal operator cannot be included in the propagation and can only be treated perturbatively. 
Hence, in Fig.~\ref{fig:phase_shifts} we plot partial-wave phase shifts both with and without the nonlocal operator included, to show the degree to which our non-perturbative calculations suffer by their exclusion. 
Even so, this scenario is much more representative of the type of calculation that would be required to include nonlocal operators at higher chiral orders (though in that case the value of the LECs are expected to be much smaller), or for direct comparison with other basis-state approaches using momentum-space interactions. 

\section{Quantum Monte Carlo Methods} \label{QMC}

QMC algorithms are routinely used to study a diverse array of physical systems, ranging from quantum chemistry and condensed matter \cite{Ceperley_1996a, Kolorenc_Mitas_2011} to cold atoms \cite{Giorgini_Pitaevskii_Stringari_2008, Gandolfi_Schmidt_Carlson_2011} and nuclear physics \cite{Gezerlis_Carlson_2010, Carlson_Gandolfi_Pederiva_etal_2015, Lynn_Tews_Carlson_etal_2016, Lynn_Tews_Carlson_etal_2017, Lynn_Tews_Gandolfi_etal_2019, Curry_Lynn_Schmidt_etal_2023, Curry_Dissanayake_Gandolfi_etal_2024, Tews_Somasundaram_Lonardoni_etal_2024}.
The QMC family of approaches is particularly attractive to the study of nuclear physics because they are \textit{ab initio} approaches that employ only controlled approximations and, in principle, solve the nuclear many-body problem exactly. 
In addition, one of the key benefits of QMC is that a given Hamiltonian can be used to study a range of diverse systems such as light- and medium-mass nuclei \cite{Lynn_Carlson_Epelbaum_etal_2014, Lynn_Tews_Carlson_etal_2017, Lonardoni_Carlson_Gandolfi_etal_2018}, neutron matter \cite{Sarsa_Fantoni_Schmidt_etal_2003, Gezerlis_Carlson_2008, Gandolfi_Gezerlis_Carlson_2015, Gandolfi_Palkanoglou_Carlson_etal_2022,Tews_Somasundaram_Lonardoni_etal_2024} that is similar to the neutron-rich matter in the outer core of neutron stars, as well as symmetric nuclear matter \cite{Gandolfi_Lovato_Carlson_etal_2014, Lonardoni_Tews_Gandolfi_etal_2020}. 
This means that QMC approaches, such as the AFDMC method employed here, can be used as a direct connection between nuclear experiments and astrophysical observations of neutron stars.

\subsection{A Brief Review of Auxiliary-Field Diffusion Monte Carlo}

The auxiliary-field diffusion Monte Carlo (AFDMC) method works by recasting the Schr\"odinger equation in imaginary time,
\begin{align} \label{Schro}
- \frac{\partial \Psi}{\partial \tau} = [\hat{H} - E_T]\Psi\,,
\end{align}
where $E_T$ is an offset introduced to aid in normalization.
By evolving a trial wavefunction in imaginary time,
the algorithm projects out the ground state of the system.
To see this, we can apply the imaginary-time evolution operator to a given trial wavefunction $\Psi_T$, which can be expanded in the exact eigenbasis of our Hamiltonian $\psi_i$, 
\begin{align}
    \Psi(\tau) &= e^{-(\hat{H}-E_T)\tau}\Psi_T \nonumber
    \\
    &= \sum_i c_i e^{-(\hat{H}-E_T)\tau}\psi_i
    \\
    &= c_0 e^{-(E_0-E_T)\tau}\psi_0 , \ \ \ \ \ \ \text{lim}\ \  \tau \rightarrow \infty\,. \nonumber
\end{align}
The excited-state terms decay to zero as $\tau$ becomes large, and by setting $E_T \approx E_0$ the ground state contribution remains finite.

For spin- and isospin-independent interactions, the imaginary time projection is carried out by considering the formal solution to Eq. (\ref{Schro}),
\begin{align}
    \braket{\bm{R}'| \Psi(\tau+\Delta \tau)} = \int d\bm{R}\ G(\bm{R}',\bm{R}; \Delta \tau) \braket{\bm{R}|\Psi(\tau)}\,,
\end{align}
where we have divided the full imaginary time into very small time steps $\tau=n\Delta \tau$, and we are propagating from $\bm{R}\rightarrow \bm{R}'$. The propagator $G(\bm{R}',\bm{R}; \Delta \tau)$ is the short-time Green's function,
\begin{align}
    G(\bm{R}',\bm{R}; \Delta \tau) = \bra{\bm{R}'}e^{-(\hat{H}-E_T)\Delta \tau}\ket{\bm{R}}\,,
\end{align}
which allows the imaginary-time evolved wave function to be written as,
\begin{align}
    \braket{\bm{R}_n | \Psi(\tau)} &= \bra{\bm{R}_n}\prod_n e^{-(\hat{H}-E_T)\Delta \tau}\ket{\Psi_T} \nonumber
    \\
    &= \int d\bm{R}_{n-1} \cdots d\bm{R}_1 d\bm{R}_0\ \label{intprop} 
    \\ 
    &\times G(\bm{R}_n,\bm{R}_{n-1};\Delta \tau) \cdots G(\bm{R}_1,\bm{R}_0;\Delta \tau) \psi_T(\bm{R}_0)\,, \nonumber 
\end{align}
where $\Psi_T(\bm{R}_0)=\braket{\bm{R}_0|\Psi_T}$.
The short-time Green's function can be approximated by using a Trotter-Suzuki approximation,
\begin{align} \label{shorttimeG}
    G(\bm{R}',\bm{R}; \Delta \tau) &= \bra{\bm{R}'}e^{-(\hat{H}-E_T)\Delta \tau}\ket{\bm{R}} \nonumber
    \\
    &\approx \bra{\bm{R}'}e^{-\Delta \tau \frac{\hat{V}_{SI}}{2}} e^{-\Delta \tau \hat{T}}e^{-\Delta \tau \frac{\hat{V}_{SI}}{2}}\ket{\bm{R}e^{E_T \Delta \tau}} \nonumber
    \\
    &= e^{-\frac{\Delta\tau}{2}[V_{SI}(\bm{R}')+V_{SI}(\bm{R})-2E_T]} \nonumber \times
    \\
    &\ \ \ \ \ \ \ \ \ \ \ \ \  \ \ \ \ \ \ \ \ \times \bra{\bm{R}'}e^{-\hat{T}\Delta\tau}\ket{\bm{R}}\,,
\end{align}
where $\hat{T}$ and $\hat{V}_{SI}$ are the kinetic and spin-independent potential operators. 
The diffusion propagator, which depends only on the kinetic energy, can be written as,
\begin{align}
    \bra{\bm{R}'}e^{-\hat{T}\Delta\tau}\ket{\bm{R}} = \biggl(\frac{m}{2\pi\hbar^2\Delta\tau}\biggl)^{\frac{3N}{2}} e^{\frac{-m(\bm{R}'-\bm{R})^2}{2\Delta \tau \hbar^2}}\,,
\end{align}
which means the full short-time Green's function for spin independent interactions is given by, 
\begin{align}
&G(\bm{R}',\bm{R};\Delta \tau) = \left(\frac{m}{2\pi\hbar^2 \Delta \tau}\right)^{\frac{3N}{2}} \text{exp}\left[\frac{-m(\bm{R}'-\bm{R})^2}{2\Delta \tau \hbar^2}\right] \nonumber
\\
&\ \ \ \ \ \ \ \ \ \ \times \text{exp}\left[-\frac{\Delta\tau}{2}[V_{SI}(\bm{R}')+V_{SI}(\bm{R})-2E_T]  \right]\,.  \label{spinI-G}
\end{align}

For spin- and isospin-dependent operators, the above procedure is less straightforward, due to the effect of the spin-isospin operators such as $\bm{\sigma}_i \cdot \bm{\sigma}_j$ , $\bm{\tau}_i \cdot \bm{\tau}_j$, etc \cite{Lynn_Tews_Gandolfi_etal_2019}. 
Historically there have been two main approaches for the handling of spin and isospin operators in QMC. 
Green's function Monte Carlo (GFMC) \cite{Carlson_1987} treats these operators by including the full spin-isospin dependence in the trial wavefunction, and explicitly summing over all spin-isospin states.
While this is an exact calculation, it suffers from an exponential scaling with the number of particles, and as a result is limited to the realm of light nuclei \cite{Pudliner_Pandharipande_Carlson_etal_1997}. 
It is worth noting that GFMC is still routinely used to study electromagnetic and electroweak observables \cite{King_Pastore_2024} as well as for few-body nuclear scattering \cite{Lynn_Tews_Carlson_etal_2016}. 

Alternatively, instead of explicitly summing over the spin-isospin states, the AFDMC approach \cite{Schmidt_Fantoni_1999} samples the spin-isospin states through rotations of the nucleon spinors. 
This is achieved through the application of a Hubbard-Stratonovich transformation, which reduces the operator dependence from quadratic to linear. 
The basic AFDMC formalism of Ref.~\cite{Schmidt_Fantoni_1999} starts by manipulating the spin-isospin dependent propagator in order to write it in a form that is amenable to a Hubbard-Stratonovich transformation. The full details on the AFDMC method, including importance sampling, spin-orbit terms, and three-body forces, can be found in the relevant literature \cite{Schmidt_Fantoni_1999,Carlson_Gandolfi_Pederiva_etal_2015, Lynn_Tews_Carlson_etal_2017, Lonardoni_Gandolfi_Lynn_etal_2018, Lonardoni_Carlson_Gandolfi_etal_2018}

In brief, AFDMC begins by considering the spin-isospin dependent local operators that appear at NLO, which also appear in the AV6 potential given by,
\begin{align}
V_{SD} &= \sum_{i<j}\biggl[ v_2\bm{\tau}_i \cdot \bm{\tau}_j + v_3\bm{\sigma}_i \cdot \bm{\sigma}_j + v_4(\bm{\tau}_i \cdot \bm{\tau}_j)(\bm{\sigma}_i \cdot \bm{\sigma}_j) \nonumber
\\
&+ v_5[3(\bm{\sigma}_i\cdot \hat{r}_{ij})(\bm{\sigma}_j\cdot \hat{r}_{ij})-\bm{\sigma}_i\cdot\bm{\sigma_j}]  \label{V6}
\\
&   + v_6[3(\bm{\sigma}_i\cdot \hat{r}_{ij})(\bm{\sigma}_j\cdot \hat{r}_{ij})(\bm{\tau}_i \cdot \bm{\tau}_j)-(\bm{\tau}_i \cdot \bm{\tau}_j)(\bm{\sigma}_i\cdot\bm{\sigma_j})] \biggl]\nonumber\,,
\end{align}
where $v_2=v_2(r_{ij})$, $v_3=v_3(r_{ij})$, etc. are radial functions multiplying the spin and isospin operators. 
After analytic manipulations, employing the well-known Hubbard-Stratonovich transformation,
\begin{align}
e^{-\frac{1}{2}\lambda \hat{O}^2} = \frac{1}{\sqrt{2\pi}}\int dx\ e^{-\frac{x^2}{2}+\sqrt{-\lambda}x\hat{O}}\,,
\end{align}
where $x$ is an auxiliary-field variable, reduces the spin and isospin dependence from quadratic to linear.
The final result is that the spin-isospin dependent interaction part of the imaginary time propagator can be written as
\begin{align} \label{SD_prop}
e^{-V_{SD}\Delta \tau} = \prod_{n=1}^{15N} \frac{1}{\sqrt{2\pi}}\int dx_n \ e^{-x_n^2/2} e^{\sqrt{-\lambda_n\Delta\tau}x_n \hat{O}_n}\,,
\end{align}
where we have 15 auxiliary fields per nucleon and $\hat{O}_n$ contains the $3N$ spin operators, the $9N$ spin-isospin operators, and the $3N$ isospin operators. 
These spin-isospin propagators of the form $e^{\bm{\sigma}\cdot \bm{n}}$, $e^{\bm{\tau}\cdot \bm{n}}$, and $e^{\bm{\sigma \tau}\cdot \bm{n}}$ are in essence rotation operators in terms of Pauli matrices which act on spinors $\bm{S}$  that describe the spin-isospin amplitudes of the nucleons in our calculation. 
The full short-time Green's function is then written as the product of Eq.~(\ref{spinI-G}) and $\bra{\bm{R}'\bm{S}'}e^{-\Delta \tau \hat{V}_{SD}/2} e^{-\Delta \tau \hat{V}_{SD}/2}\ket{\bm{R}\bm{S}}e^{E_T \Delta \tau}$, 
and we are now able to straightforwardly handle spin- and isospin-dependent interactions.

\subsection{Trial Wave Function}

One of the key characteristics of any QMC calculation is the use of a physics-informed trial wave function.
The general trial wave function used in our calculations is of the form \cite{Lonardoni_Gandolfi_Lynn_etal_2018},
\begin{align}
    \Psi(\bm{R},\bm{S})&=\braket{\bm{R},\bm{S}|\Psi_T}
    \\
    &=\bra{\bm{R},\bm{S}} \biggl[ \prod_{i<j} f_c(r_{ij}) \biggl] \biggl[ 1 + \sum_{i<j,p}f_p(r_{ij}) O_{ij}^p \biggl] \ket{\Phi}\,,\nonumber
\end{align}
where the $\ket{\bm{R},\bm{S}}$ capture the spatial and spin-isospin coordinates and the $p$ sum is over the five operators that appear in Eq.~(\ref{V6}). 
The pair correlation functions $f(r_{ij})$ are calculated by solving a one-body Schr\"odinger equation in terms of the separation between pairs \cite{Pandharipande_Wiringa_1979, Carlson_Gandolfi_Pederiva_etal_2015}. 
The $\ket{\Phi}$ can take on different forms depending on the system we are interested in. 
For our deuteron calculations, we employ a shell-model like wave function developed for the investigation of light nuclei \cite{Lonardoni_Gandolfi_Lynn_etal_2018} that employs a sum of Slater determinants consisting of single-particle orbitals multiplied with variational parameters. 
These parameters, and similar ones in the neutron-matter wavefunction, can be optimized at the VMC level \cite{Sorella_2001} before carrying out the full AFDMC calculation.
To study neutron matter, the $\ket{\Phi}$ is built up using plane-wave orbitals.
To avoid computational complexity, we also employ a simplified version of spin-orbit correlations following Ref.~\cite{Lonardoni_Tews_Gandolfi_etal_2020}.

\section{Second-Order Perturbation Theory in Quantum Monte Carlo}

Consider a Hamiltonian consisting of a generic unperturbed Hamiltonian, $H_0$, and a perturbing two-body potential $V^{\prime}$,
\begin{equation}
    H = H_0 + \lambda V^{\prime}\,,
\end{equation}
where $\lambda$ is a small dimensionless parameter.
Stationary perturbation theory allows one to write the energies and eigenstates of the perturbed system in terms of corrections to the energies and eigenstates of the unperturbed system, 
\begin{align}
    E_n &= E_n^{(0)} + \lambda E_n^{(1)} + \lambda^2 E_n^{(2)} + \cdots\,,
    \\
    \ket{\phi_n} &= \ket{\phi_n^{(0)}} + \lambda \ket{\phi_n^{(1)}} + \lambda^2 \ket{\phi_n^{(2)}} + \cdots\,,
\end{align}
where $E_n^{(i)}$ is the $i^{\text{th}}$ order correction to the unperturbed $n^{\text{th}}$ energy eigenvalue. 
The first- and second-order corrections to the ground-state energy of the unperturbed system $H_0$ are then written as,
\begin{align}
E_0^{(1)}&=\langle\phi_0^{(0)}|V^{\prime}|\phi_0^{(0)}\rangle \label{1storder}\,,
\\
E_0^{(2)}&=-\sum_{k\neq0} \frac{|\langle\phi_k^{(0)}|V^{\prime}|\phi_0^{(0)}\rangle|^2}{E_k^{(0)}-E_0^{(0)}}\,. \label{2ndOrder}
\end{align}
Since the eigenstates and energies considered in this work are many-body entities, these perturbative corrections cannot be evaluated analytically. 
Due to the fact the AFDMC method described above (and any projector QMC approach) is primarily concerned with projecting out only the ground state energy and wave function, Eq.~(\ref{1storder}) is straightforward to compute, while the calculation of Eq.~(\ref{2ndOrder}) is highly non-trivial due to our lack of knowledge regarding the full energy spectrum $E_k^{(0)}$. 
Recent work has developed an approach for calculating the second-order energy correction in a QMC context~\cite{Lu_Li_Elhatisari_etal_2022, Curry_Lynn_Schmidt_etal_2023}. 
In this work, we follow and review the formalism of Ref.~\cite{Curry_Lynn_Schmidt_etal_2023} before applying it to a perturbation containing nonlocal terms as discussed in Sec.~\ref{NuclearH}.

As AFDMC projects out the ground state of our system, we assume we have access to the ground state $|\phi_0^{(0)}\rangle$, and consider,
\begin{align} \label{CompleteI}
I(\mathcal{T})=\int_0^{\mathcal{T}}d\tau \left< \phi_0^{(0)} \bigg| V^{\prime} e^{-\left[H_0-E_0^{(0)}\right]\tau}V^{\prime} \bigg| \phi_0^{(0)} \right>\,,
\end{align}
where $e^{-[H_0 - E_0^{(0)}]\tau}$ is the imaginary-time propagator for the unperturbed system. By an insertion of a resolution of the identity for the unperturbed system, we find
\begin{align}
&I(\mathcal{T}) =
\\
&\sum_{k=0}^{\infty}\int_0^{\mathcal{T}} \hspace{-6pt} d\tau e^{-
\left[E_k^{(0)}-E_0^{(0)}\right]\tau} \left<\phi_0^{(0)}| V^{\prime} | \phi_k^{(0)} \right> \left<\phi_k^{(0)}| V^{\prime} | \phi_0^{(0)} \right>\,. \nonumber
\end{align}
The sum is then split into a term for $k=0$ and a term for $k\neq 0$,
\begin{align} \label{split}
I(\mathcal{T})&=\int_0^{\mathcal{T}} d\tau |\langle\phi_0^{(0)}|V^{\prime}|\phi_0^{(0)}\rangle|^2 + \nonumber
\\
&+ \sum_{k\neq0} \int_0^{\mathcal{T}} d\tau e^{-\left[E_k^{(0)}-E_0^{(0)}\right]\tau}|\langle\phi_k^{(0)}|V^{\prime}|\phi_0^{(0)}\rangle|^2\,.
\end{align}
We next integrate over imaginary time. 
The first term acquires a factor of $\mathcal{T}$, and the second term gives, 
\begin{align}
&\sum_{k\neq0} \int_0^{\mathcal{T}} d\tau e^{-\left[E_k^{(0)}-E_0^{(0)}\right]\tau}|\langle\phi_k^{(0)}|V^{\prime}|\phi_0^{(0)}\rangle|^2 \nonumber
\\
&= \sum_{k\neq0}^{\infty} |\langle\phi_k^{(0)}|V^{\prime}|\phi_0^{(0)}\rangle|^2 \int_0^{\mathcal{T}} d\tau e^{-\left[E_k^{(0)}-E_0^{(0)}\right]\tau} \nonumber
\\
&= - \sum_{k\neq0}^{\infty} |\langle\phi_k^{(0)}|V^{\prime}|\phi_0^{(0)}\rangle|^2 \times \frac{1}{E_k^{(0)}-E_0^{(0)}}  e^{-[E_k^{(0)}-E_0^{(0)}]\tau} \bigg|_0^\mathcal{T} \label{fitform} \nonumber
\\
&= - \sum_{k \neq 0}^{\infty} \frac{\left |\bra{\psi_k^{(0)}}V^{\prime}\ket{\psi_0^{(0)}} \right|^2}{E_k^{(0)} - E_0^{(0)}} \left[e^{-\left[E_k^{(0)}-E_0^{(0)}\right]\mathcal{T}} - 1 \right]\,. 
\end{align}
Finally, because we know $E_k^{(0)}-E_0^{(0)} > 0$, the exponential term vanishes in the limit of long imaginary time to give
\begin{align}
    \hspace{-5pt}I(\mathcal{T})= |\langle\phi_0^{(0)}|V^{\prime}|\phi_0^{(0)}\rangle|^2 \mathcal{T} +  \sum_{k \neq 0}^{\infty} \frac{\left |\bra{\psi_k^{(0)}}V^{\prime}\ket{\psi_0^{(0)}} \right|^2}{E_k^{(0)} - E_0^{(0)}}\,,
\end{align}
which by comparison with Eqs. (\ref{1storder}) and (\ref{2ndOrder}) is equal to
\begin{align}
    I(\tau \rightarrow \infty) = (E_0^{(1)})^2 \tau - E_0^{(2)}.
\end{align}
Therefore, by an implementation of Eq.~(\ref{CompleteI}) one is able to calculate both the first- and second-order perturbative corrections to the ground-state energy in a QMC calculation. 
This provides not only our quantity of interest, $E_0^{(2)}$, but also a reliable method to check our calculations through a secondary means of calculating $E_0^{(1)}$.

In an AFDMC calculation, the full imaginary time is divided into many small-$\tau$ segments, and so Eq. (\ref{CompleteI}) is rewritten as,
\begin{align}
    I(\mathcal{T})= \braket{\psi_0^{(0)} \biggl| V^{\prime}\prod_n e^{- \left[\hat{H}_0-E_0 \right] \Delta \tau} V^{\prime} \biggl| \psi_0^{(0)}}\,. \label{Istep}
\end{align}
We can also consider a single step of the Monte Carlo integral,
\begin{align} \label{I_qmc}
    I_n = \hspace{-0.15cm}\int d\bm{R}' d\bm{R}\ \hat{V}^{\prime}\psi_T(\bm{R}') G \cdot \hat{V}^{\prime} \psi_T(\bm{R})\,, 
\end{align}
stepping from $\bm{R}$ to $\bm{R}'$, where $G = G(\bm{R}',\bm{R},\bm{S}',\bm{S};\Delta \tau)$ is the short-time Green's function propagator as discussed in Section \ref{QMC}.
We have suppressed the $\bm{S}$ notation in the remainder of this section to aid in legibility, all $\bm{R}$ and $\bm{R}'$ have accompanying $\bm{S}$ and $\bm{S}'$.

We then carry out an importance-sampling scheme similar to what is commonly used in DMC \cite{Foulkes_Mitas_Needs_etal_2001} and AFDMC \cite{Lonardoni_Gandolfi_Lynn_etal_2018}, with the modification that we are now considering two $V^{\prime}$ operators separated by a diffusion step.
It is worth a reminder that our perturbation $\hat{V}^{\prime}$ is no longer assumed to be a scalar operator as in Ref.~\cite{Curry_Lynn_Schmidt_etal_2023}. 
We start by multiplying and dividing by the trial wave function for each configuration to find
\begin{align}
    I_n \hspace{-2pt}&= \hspace{-5pt}\int d\bm{R}' d\bm{R}\  \frac{\psi_T(\bm{R}')}{\psi_T(\bm{R}')} \hat{V}^{\prime}\psi_T(\bm{R}') G \frac{\psi_T(\bm{R})}{\psi_T(\bm{R})} \hat{V}^{\prime} \psi_T(\bm{R}) \nonumber
    \\
    \hspace{-2pt}&= \hspace{-5pt}\int d\bm{R}' d\bm{R} \left[ \frac{\hat{V}^{\prime} \psi_T(\bm{R}')}{\psi_T(\bm{R}')} \right] \left[\frac{\psi_T(\bm{R}') G}{\psi_T(\bm{R})} \right]\psi_T(\bm{R}) \hat{V}^{\prime} \psi_T(\bm{R}) \nonumber
    \\
    \hspace{-2pt}&= \hspace{-5pt}\int d\bm{R}' d\bm{R}\ V_L^{\prime}(\bm{R}') \tilde{G}\ \psi_T(\bm{R}) \hat{V}^{\prime} \psi_T(\bm{R})\,,
\end{align}
where $\tilde{G}$ is the importance sampled short-time propagator \cite{Pudliner_Pandharipande_Carlson_etal_1997} and we have introduced a ``local energy" version of our nonlocal perturbation, $V_L^{\prime}(\bm{R}')= \frac{\hat{V}^{\prime} \psi_T(\bm{R}')}{\psi_T(\bm{R}')}$. 
To show that the formalism from \cite{Curry_Lynn_Schmidt_etal_2023} extends to nonlocal operators, we carry out a final importance-sampling procedure for the $\bm{R}$ configuration to give
\begin{align}
    I_n &= \int d\bm{R}' d\bm{R}\ V_L^{\prime}(\bm{R}') \tilde{G}\ \frac{\psi_T(\bm{R})}{\psi_T(\bm{R})}\psi_T(\bm{R}) \hat{V}^{\prime} \psi_T(\bm{R}) \nonumber
    \\
    &= \int d\bm{R}' d\bm{R}\ V_L^{\prime}(\bm{R}') \tilde{G}\ \psi_T^2(\bm{R}) V_L^{\prime}(\bm{R})\,,
\end{align}
which can be computed as an importance-sampled Monte Carlo integral,
\begin{align} 
    I_n \approx \frac{\sum_i^{\mathcal{N}} w_iV_L^{\prime}(\bm{R}_i')V_L^{\prime}(\bm{R}_i)}{\sum_i^{\mathcal{N}}w_i}\,, \label{Istep_mc}
\end{align}
where $\mathcal{N}$ is the number of walkers and the weights $w_i$ are calculated according to the branching component of Eq. (\ref{shorttimeG}). 
The $\bm{R}'$ are sampled from $\tilde{G}$ through our diffusion steps, and the $\bm{R}$ are sampled from $\psi_T^2$. 
Note that if $\hat{V}^{\prime}$ were a scalar operator, this formalism reduces exactly to the result from \cite{Curry_Lynn_Schmidt_etal_2023}. 

\begin{figure}[t]
\centering
    \includegraphics[width=0.45\textwidth]{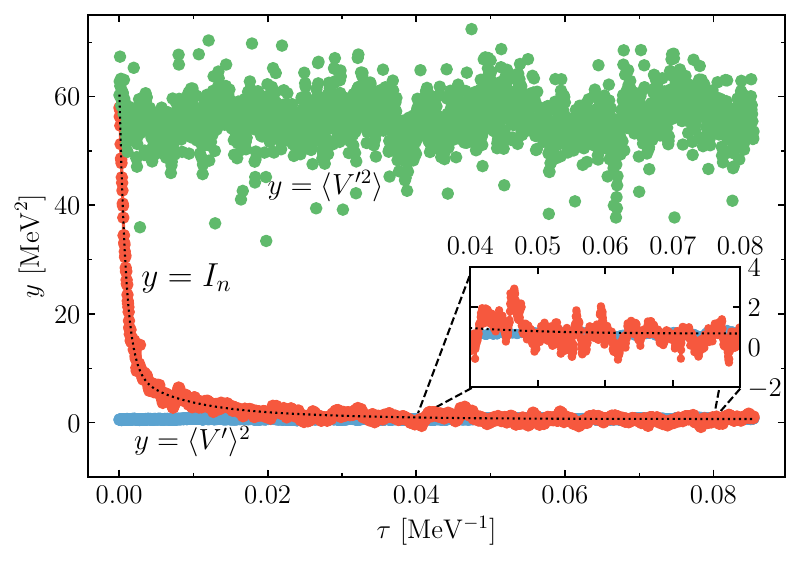}
   \caption{An AFDMC calculation of Eq.~(\ref{Istep}) for the ${k^2 \bm{\sigma}_1 \cdot \bm{\sigma}_2}$ operator for the deuteron system, marked with red points and labeled $I_n$. 
   By comparing with Eq.~(\ref{I_qmc}) we see the expected behavior. At very small $\tau$, the two evaluations of $\hat{V}'$ are separated by only a very small $\Delta \tau$ step, and so we expect $I_n \approx \left< V^{\prime 2} \right>$ (green points). Similarly our calculations match the expectation that in the limit of long $\tau$, $I_n$ should decay to $\left< V^{\prime} \right>^2$ (blue points) based on the form of Eq.~(\ref{split}). Though the calculation is dramatically more variable than what is seen for local operators (see Fig. \ref{fig:ext_est}), it is still straightforward to fit and extract the perturbative corrections.}
\label{fig:example_calc}
\end{figure}
\begin{figure}[t]
\centering
\includegraphics[width=0.5\textwidth]{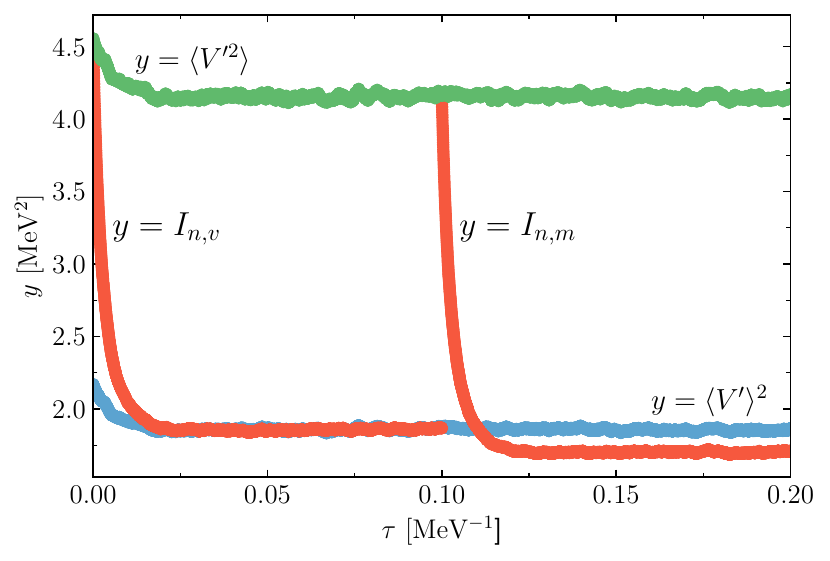} 
   \caption{Extrapolated-estimate calculation for a perturbative correction to the deuteron binding energy using a local Gaussian perturbation for illustrative purposes only \cite{Curry_Lynn_Schmidt_etal_2023} (same color scheme as Fig. (\ref{fig:example_calc})). As discussed in the text, the second-order correction is extracted from two AFDMC calculations labeled by $I_{n,v}$ and $I_{n,m}$ respectively. The initial ``variational" AFDMC calculation, $I_{n,v}$, starts at $\tau=0$ and the second ``mixed" AFDMC calculation, $I_{n,m}$, is started once the walkers have reached the ground-state configuration. The mixed and variational second-order corrections are found from fitting to these two curves.}
\label{fig:ext_est}
\end{figure}
\begin{figure*}
\includegraphics[scale=0.55]{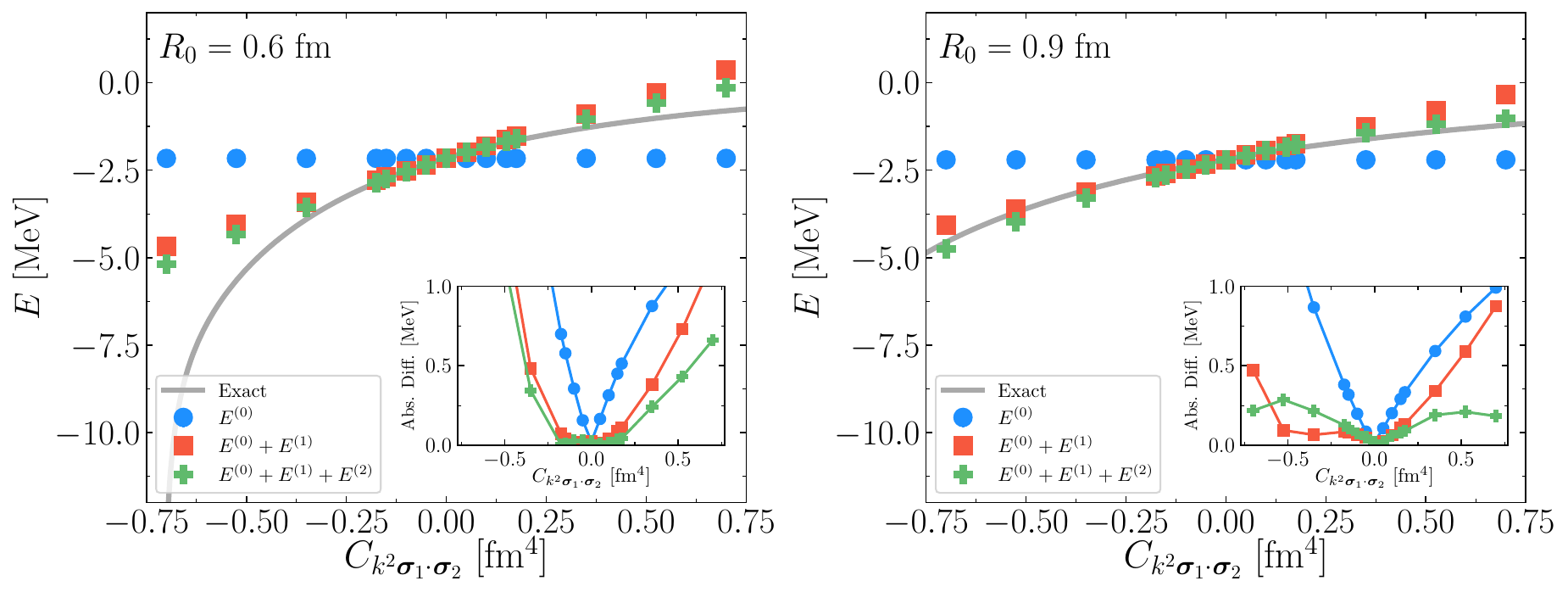} 
   \caption{AFDMC calculations including perturbative corrections for the deuteron compared against exact solutions of the Lippmann-Schwinger equations (grey lines).
   These results are obtained using our ``frozen" NLO$_0$ interaction, where only the local LECs are included in the fit to phase-shift data, and the value of the nonlocal LEC can be varied by hand. Our AFDMC results include terms up to zeroth order (blue circles), first order (red squares), and second order (green diamonds). These calculations show that the perturbation-theory formalism from \cite{Curry_Lynn_Schmidt_etal_2023} is applicable to realistic nuclear interactions and the more sophisticated AFDMC algorithm. 
   }
\label{fig:deuteron_frozen}
\end{figure*}

An example calculation of Eq. (\ref{Istep}) for a nonlocal operator is shown in Fig. \ref{fig:example_calc}, red points labeled $I_n$. 
The first- and second-order perturbative corrections are extracted from this calculation by carrying out a fit that is
informed by the expected integrand of Eq. (\ref{fitform}). 
In all practical calculations, one does not have access to the true ground state of the system and so the above prescription must be modified to account for this. As is common in QMC calculations, we employ a form of the extrapolated estimate, 
\begin{align}
    \braket{O}_{\text{ext}} = 2\braket{O}_{\text{m}} - \braket{O}_{\text{v}},
\end{align}
where the subscript $m$ refers to a mixed estimate and $v$ refers to a variational estimate. In Fig. \ref{fig:ext_est}, we consider a local Gaussian perturbation, $V'(r) = e^{-(\bm{r}_2 - \rm{r}_1)^2}$, acting between two particles, in order to illustrate the nature of the extrapolated estimate.

To find our ``variational" estimate, we perform a standard AFDMC calculation, where our walkers are initially distributed according to $\psi_T^2$ by an initial VMC run.
This allows us to compute a variational estimate of Eq.~(\ref{Istep}), where the initial and propagated walkers are distributed according to
\begin{align}
\bm{R}_{i,v} &\longrightarrow \psi_T^2 \nonumber
\\
\bm{R}_{i,v}' &\longrightarrow \tilde{G}(\bm{R}',\bm{R};\Delta \tau)= \psi_T(\bm{R}')\frac{G(\bm{R}',\bm{R};\Delta \tau)}{\psi_T(\bm{R})}\,.
\end{align}

To carry out our mixed estimate, we carry out a second AFDMC calculation, not starting from walkers distributed according to $\psi_T^2$, but from the end of the previous AFDMC run, with the walkers distributed according to $\psi_0^2$. 
While tempting to think that this mixed estimate should be equivalent to the pure estimate with $\psi_T = \psi_0$, we can see that this is not the case by looking at how the initial and propagated walkers are distributed,
\begin{align}
\bm{R}_{i,m} &\longrightarrow \psi_0^2 \nonumber
\\
\bm{R}_{i,m}' &\longrightarrow \tilde{G}(\bm{R}',\bm{R};\Delta \tau)= \psi_T(\bm{R}')\frac{G(\bm{R}',\bm{R};\Delta \tau)}{\psi_T(\bm{R})}\,,
\end{align}
as well as Fig.~\ref{fig:ext_est}.
Even though it is true that the $\bm{R}$ configurations are already distributed according to the ground state, our $\bm{R}'$ configurations still have trial wave function dependence, and so this is properly thought of as a mixed estimate. 
As the initial trial wave function approaches the true ground-state wave function, whether by including relevant physics or by optimization, the difference between the long-$\tau$ behavior of these two curves will decrease until the point where they would overlap if $\psi_T=\psi_0$, in which case an extrapolated estimate would be unnecessary.
For the remainder of this work, we will drop the subscripts on the ground-state energy and its corrections, as we are not currently investigating any excited states. 

\section{Deuteron} \label{deuteron}
The deuteron holds a special significance in nuclear physics as the only two-nucleon bound state that exists in nature.
The binding energy of the deuteron can be found exactly by solving the Lippmann-Schwinger equation, which means it often serves as a testing ground for new calculations and theoretical approaches.

As an initial test case for applying our second-order perturbation theory formalism,  
we consider the ``frozen" NLO$_0$ interaction, where we have used only the local LECs when fitting to phase-shift data.
This allows us to then vary the nonlocal LEC over a range of values and test how well our perturbation theory performs. 
One would expect perturbation theory to excel in regions with small perturbations, i.e., small values of the nonlocal LEC, and perhaps struggle when the LEC becomes large. 

Our results are plotted in Fig. \ref{fig:deuteron_frozen} for two choices of the coordinate space cutoff $R_0$. 
We have carried out these calculations for both a very hard cutoff ($R_0 = 0.6\ \text{fm}$) as well as a very soft cutoff ($R_0 = 0.9\ \text{fm}$). 
What we call $C_{k^2 \bm{\sigma}_1\cdot \bm{\sigma}_2}$ here and below was denoted by $\tilde{C_3}$ in Eq. (\ref{coordinatespace}). In this plot, and all following plots, we compare the exact/expected result in grey, to our AFDMC results which include terms up to zeroth order (blue circles), first order (red squares) and second order (green diamonds).
\begin{figure}[t]
\centering
\includegraphics[width=0.47\textwidth]{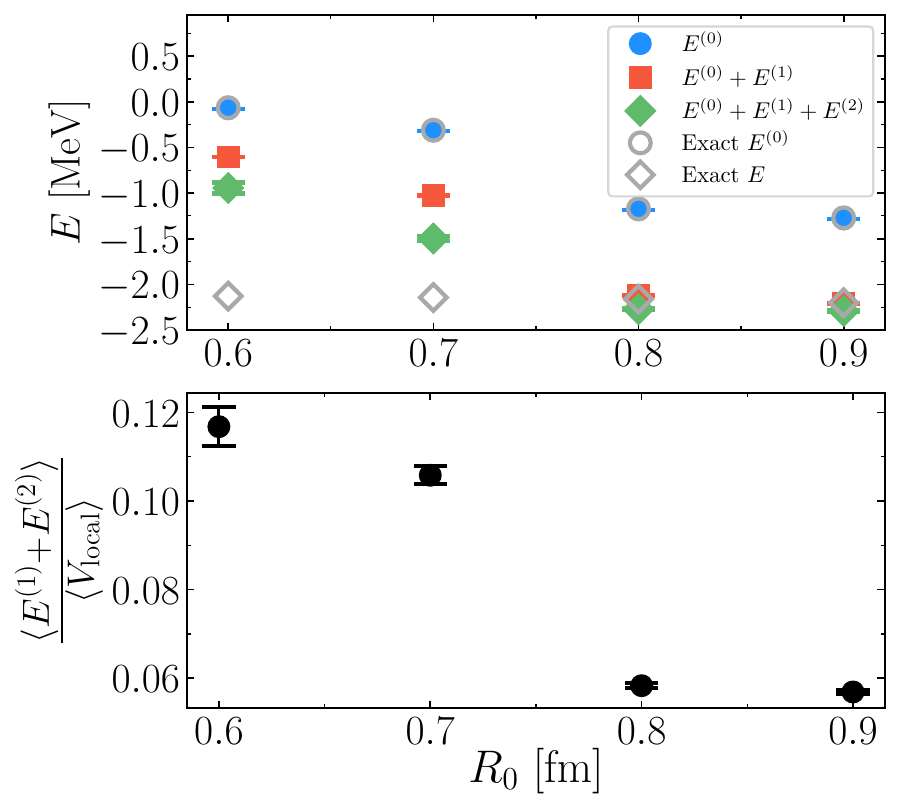} 
   \caption{AFDMC calculations including perturbative corrections for the deuteron compared against the exact Lippmann-Schwinger solution. These results are obtained using our ``refit" NLO$_{\text{I}}$ interaction, where all seven LECs are included in the fit to phase-shift data, though the nonlocal $k^2 \bm{\sigma}_1 \cdot \bm{\sigma}_2$ operator is only included perturbatively in the full many-body calculations. 
   The upper panel shows the order-by-order AFDMC results, while the lower panel shows the ratio of the perturbative contribution to the propagated local potential contribution, as a measure of pertrubativeness (details in text).
   }
\label{fig:deuteron_refit}
\end{figure}
\begin{figure} [t]
\centering
\includegraphics[width=0.48\textwidth]{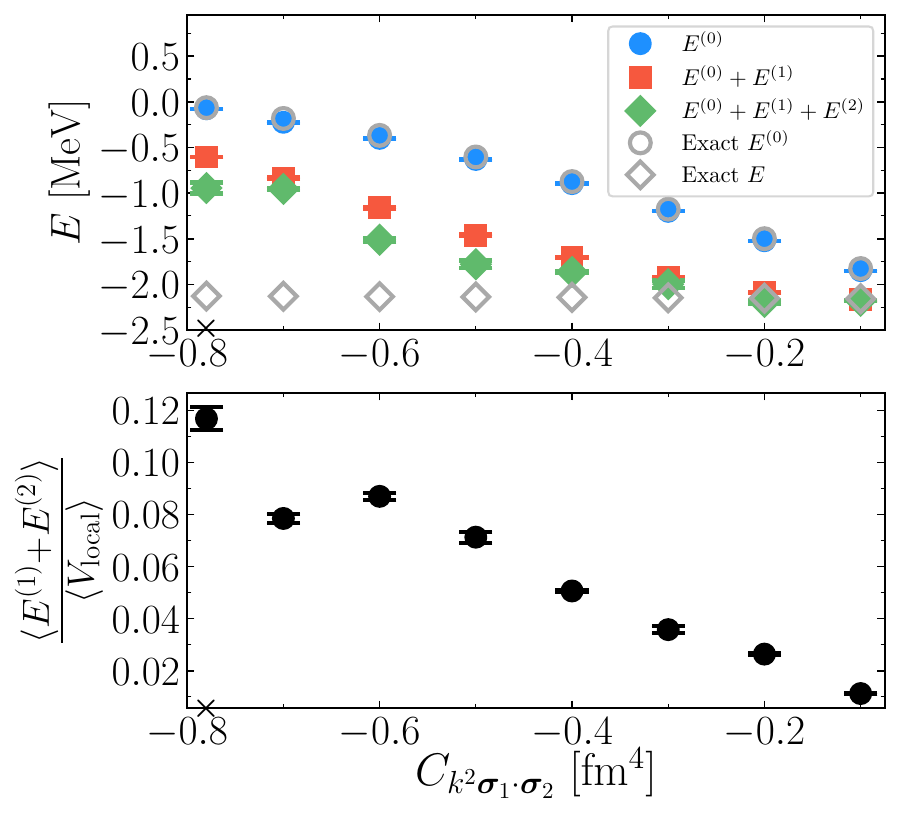} 
   \caption{AFDMC calculations similar to those in Fig.~\ref{fig:deuteron_refit}, where we have instead used our ``constrained"  NLO$_{\text{II}}$ interactions at $R_0=0.6\ \text{fm}$, where all seven LECs are included in the fit to phase-shift data, but the nonlocal LEC is bounded and only included in the full many-body calculations perturbatively. 
   The upper panel shows the order-by-order AFDMC results, while the lower panel shows the ratio of the perturbative contribution to the propagated local potential contribution. The results at the far left of this plot (denoted with an $\times$ on the x-axis) correspond to the $R_0=0.6\ \text{fm}$ points from Fig. \ref{fig:deuteron_refit}.}
\label{fig:deuteron_varied}
\end{figure}
We can see that for both cutoffs, the perturbative results agree perfectly with the exact results in the region of small LEC values.
When studying larger LEC values we find that in general the softer interaction is much more perturbative, with the second-order results being in good agreement with the exact solutions. 
For some negative values of the LEC in the softer interaction, the second-order correction overshoots the analytic result.
However, the perturbative calculation seems to converge. 

The calculations using NLO$_0$ in Fig. \ref{fig:deuteron_frozen} are a test case, and in practical calculations the nonlocal operator must be included into the phase-shift fits in order to have a complete set of operators for the NLO chiral interactions. 
Therefore, as seen in Fig. \ref{fig:phase_shifts} and discussed in section \ref{NuclearH}, we have developed a ``refit" NLO$_{\text{I}}$ potential by refitting the local LECs while including the nonlocal term, over a range of coordinate-space cutoffs. 
However, since the nonlocal operator cannot be included in the imaginary-time propagation, we can only expect to recover the correct physics by including it in our many-body calculation perturbatively.
In contrast to the previous case, there is no \textit{a priori} reason to expect that the nonlocal correction will be perturbative. 

As the results in Fig. \ref{fig:deuteron_refit} show, we find that the perturbativeness of the nonlocal contribution to the NLO Hamiltonian 
seems to depend strongly on the coordinate-space cutoff used in the interaction. 
For softer interactions, i.e., larger values of $R_0$, our perturbative AFDMC results do a very good job matching the exact solutions, while for harder interactions the second-order corrections are insufficient. 
While they move the results closer to the exact solution, there is little to no evidence of convergence. 

However, the two sets of calculations for the deuteron presented in Figs.~\ref{fig:deuteron_frozen} and~\ref{fig:deuteron_refit} illustrate a way to perform calculations with nonlocal interactions also at smaller $R_0$.
Although Fig. \ref{fig:deuteron_refit} shows that nonlocal operators are difficult to include perturbatively for hard cutoff interactions, Fig. \ref{fig:deuteron_frozen} shows that for small LEC values the perturbative calculations converge also for such cutoffs. 
Therefore, in the following we bound the size of the nonlocal LEC to be small when fitting to phase-shift data.
As a consequence, this ensures the nonlocal operator to be perturbative. 
To test this, we introduce a new ``constrained" NLO$_{\text{II}}$ interaction, where we have repeated the fit to phase shift procedure for a hard cutoff NLO interaction $(R_0=0.6\ \text{fm})$, and have constrained the strength of the nonlocal LEC to be within some bound, as shown in the x-axis of Fig.~\ref{fig:deuteron_varied}. 

We show in Fig.~\ref{fig:deuteron_varied} that there is a region where the perturbative approach is justified, and we can be reasonably confident in treating the nonlocal operator perturbatively. 
We have identified the ratio of the perturbative contributions to the potential energy and the propagated local potential energy, $\braket{E^{(1)}+E^{(2)}}/\braket{V_{\text{local}}}$, as a good measure of perturbativeness, as shown in the lower panels of Figs.~\ref{fig:deuteron_refit} and~\ref{fig:deuteron_varied}. 
We find that for small values of this measure,
\begin{align} \label{perturbativeness}
    \frac{\braket{E^{(1)}+E^{(2)}}}{\braket{V_{\text{local}}}} << 1\,,
\end{align}
there is a reasonable expectation of convergence in the perturbative calculations, which could be further tested through the development of third-order perturbative corrections in QMC calculations.

\section{Neutron Matter}

\begin{figure*}
\centering
\includegraphics[width=1\textwidth]{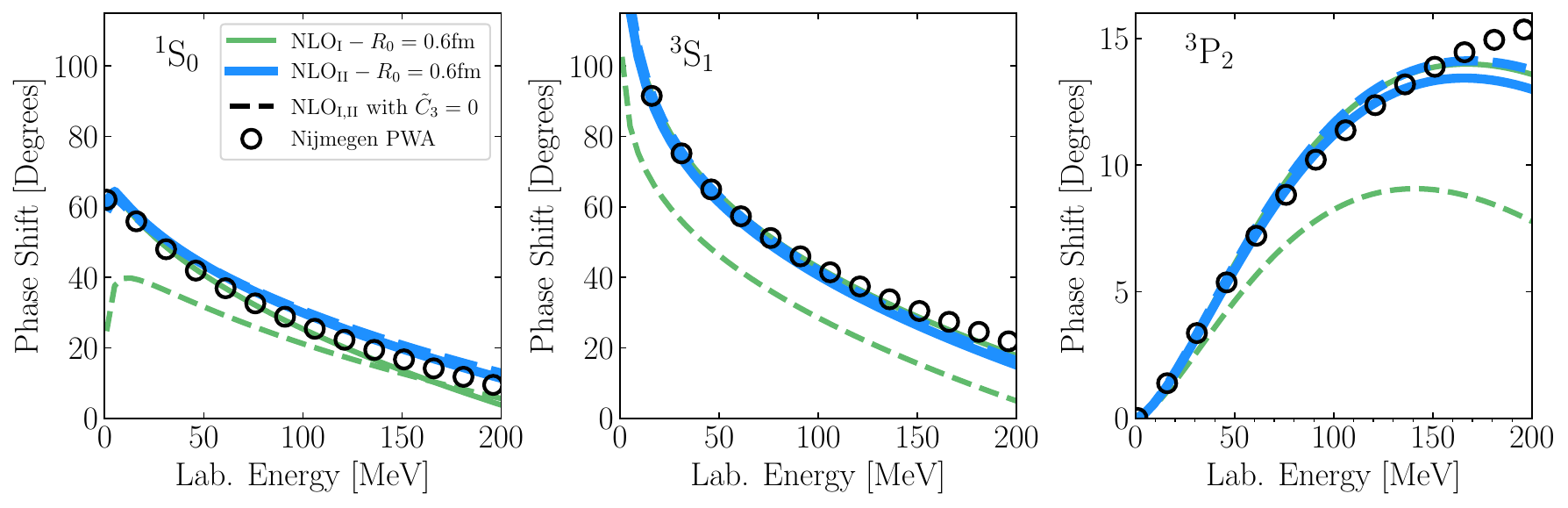} 
   \caption{Selected partial-wave phase shifts similar  to Fig. \ref{fig:phase_shifts}, for the $R_0=0.6\ \text{fm}$ hard cutoff using both NLO$_{\text{I}}$ and NLO$_{\text{II}}$ (with $C_{k^2 \bm{\sigma}_1 \cdot \bm{\sigma}_2} = -0.1\ \text{fm}^4$). 
   The solid lines describe the results of our fit to phase shifts when including the nonlocal operator, and the dashed lines show results when the nonlocal LEC has been set to zero. 
   The difference between the solid and dashed lines is a guide to the physics missing from our non-perturbative many-body calculations.} 
\label{fig:constrained_phaseshifts}
\end{figure*}

Previous work \cite{Lynn_Carlson_Epelbaum_etal_2014, Lynn_Tews_Carlson_etal_2017, Curry_Lynn_Schmidt_etal_2023} investigating both light nuclei and neutron matter struggled to treat hard cutoff interactions perturbatively. 
Constraining the nonlocal LEC to a perturbative value alleviates some of the problems. 
In order to further test this approach, we use our ``constrained" NLO$_{\text{II}}$ interaction, corresponding to $C_{k^2 \bm{\sigma}\cdot \bm{\sigma}}=-0.1\, \text{fm}^4$, for a calculation of the neutron matter equation of state.
First, we compare the partial-wave phase shifts similarly to Fig. \ref{fig:phase_shifts} for both NLO$_{\text{I}}$ and NLO$_{\text{II}}$, see Fig. \ref{fig:constrained_phaseshifts}.
We find  that in the case of the ``constrained" NLO$_{\text{II}}$, the difference between the fits that include the nonlocal operator and the phase shifts calculated after setting the nonlocal LEC to zero is negligible, as one would expect for a perturbative contribution.

\begin{figure} [b]
\centering
\includegraphics[width=0.48\textwidth]{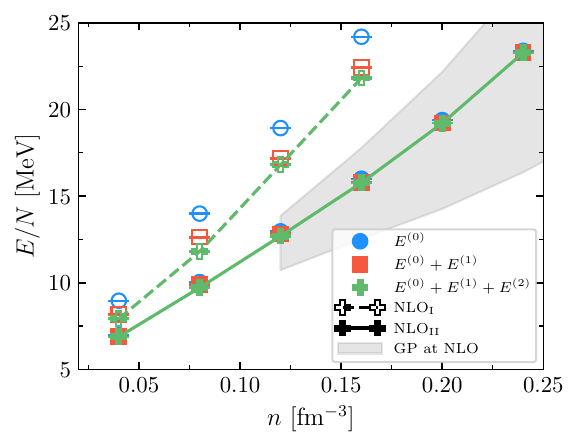} 
   \caption{AFDMC calculations of the pure neutron matter equation of state as function of number density with hard cutoff $(R_0 = 0.6\ \text{fm})$ interactions.  
   We include results for two NLO interactions, the ``refit" NLO$_{\text{I}}$ and the ``constrained" NLO$_{\text{II}}$, whose details are discussed in the main text. 
   The grey uncertainty band describes the NLO results of Ref.~\cite{Tews_Somasundaram_Lonardoni_etal_2024} estimating uncertainties using the prescription of Ref.~\cite{Drischler_Furnstahl_Melendez_etal_2020}. 
   We find that by constraining the strength of the nonlocal LEC we are able to find good agreement with these chiral EFT uncertainty estimates.}
\label{fig:eos_neutrons}
\end{figure}

To model neutron matter, we simulate a system of 14 neutrons in a box with  periodic boundary conditions at various neutron densities $(0.04\ \text{fm}^{-3} -- 0.24\ \text{fm}^{-3})$, probing from the low-density regime to beyond nuclear saturation density. 
The choice of 14 neutrons is motivated by previous studies into the finite-size effects of neutrons in periodic boundary conditions \cite{Gandolfi_Illarionov_Schmidt_etal_2009, Sarsa_Fantoni_Schmidt_etal_2003}, providing a reasonable compromise between describing the thermodynamic limit and 
computational efficiency. 
We compare our perturbative AFDMC calculations against recent results at NLO with similar regulator scheme and scale~\cite{Tews_Somasundaram_Lonardoni_etal_2024} (for 66 neutrons), that include uncertainty estimates using a Gaussian process approach \cite{Drischler_Furnstahl_Melendez_etal_2020}. 
The comparison between our 14 neutron AFDMC calculations and 66 neutron chiral EFT uncertainty bands is justified due to negligible finite-size effects as shown in \cite{Gezerlis_Tews_Epelbaum_etal_2013}.

We show the results of our calculations with NLO$_{\text{I}}$ and NLO$_{\text{II}}$ in Fig. \ref{fig:eos_neutrons}.
For NLO$_{\text{I}}$ with a sizable nonlocal contribution, the perturbative inclusion fails, and the neutron matter equation of state falls well outside the chiral EFT uncertainty bands. 
However, when using  NLO$_{\text{II}}$, the neutron matter equation of state agrees very well with the chiral EFT uncertainty bands, even though we the nonlocal operator is included perturbatively. 

These calculations serve as evidence that nonlocal operators can be included perturbatively even for hard cutoff interactions, provided that care is taken in order to ensure their perturbativeness (which can be quantified using the measure of perturbativeness introduced in Eq.~(\ref{perturbativeness})). 
Since we have shown that nonlocal operators can be included perturbatively and controlled appropriately, this work opens the door for future calculations of the neutron matter EOS using chiral EFT interactions including terms up to N$^3$LO and beyond, where the nonlocal terms cannot be avoided but are expected to be small.
These future calculations making use of our perturbative framework will provide greater constraints on the neutron matter EOS, with exciting implications for the structure of neutron stars. 

\section{Conclusion}

In summary, we have extended the formalism of second-order perturbative calculations to the AFDMC many-body method, and investigated how nonlocal operators can be included perturbatively in AFDMC calculations in a self-consistent approach. 
After studying the deuteron with interactions that are designed to be perturbative, we carry out realistic AFDMC calculations where a nonlocal operator is included perturbatively for both the deuteron and the neutron matter equation of state. 
We find considerable evidence of cutoff-dependent perturbativeness, and also develop a reasonable approach to constrain nonlocalities such that they can be included perturbatively even for hard-cutoff interactions. 
We have also introduced a measure of perturbativeness for a given nonlocal contribution, which holds promise of being widely applicable in many-body calculations, and serves as a guiding principle for future QMC calculations employing interactions at N$^3$LO. 
The ability to include nonlocal operators perturbatively should also allow for greater comparison between coordinate space QMC calculations and other basis-state many-body methods that employ chiral interactions. 
Finally, our calculations serve as both a proof of principle and a strong starting point for including nonlocal operators at higher chiral orders perturbatively in an AFDMC calculation.  

\section{Acknowledgements}
The authors are grateful to K.E. Schmidt, J.E. Lynn, J. Carlson, B.T. Reed, and G. Palkanoglou for insightful discussions, and to H. G\"ottling for providing the NLO uncertainty band in Fig.~\ref{fig:eos_neutrons}.
The work of R.C. and A.G. was supported by the Natural Sciences and Engineering Research Council (NSERC) of Canada and the Canada Foundation for Innovation (CFI). 
I.T. and S.G. were supported by the U.S. Department of Energy, Office of Science, Office of Nuclear Physics, under contract No. DE-AC52-06NA25396, and by the U.S. Department of Energy, Office of Science, Office of Advanced Scientific Computing Research, Scientific Discovery through Advanced Computing (SciDAC) NUCLEI program. 
R.S. acknowledges support from the Nuclear Physics from Multi-Messenger Mergers (NP3M) Focused Research Hub which is funded by the National Science Foundation under Grant Number 21-16686. 
R.C., R.S., and I.T. were also supported by the Laboratory Directed Research and Development program of Los Alamos National Laboratory under project number 20220541ECR. 
Computational resources have been provided by Compute Ontario through the Digital Research Alliance of Canada, the Los Alamos National Laboratory Institutional Computing Program, which is supported by the U.S. Department of Energy National Nuclear Security Administration under Contract No. 89233218CNA000001, and by the National Energy Research Scientific Computing Center (NERSC), which is supported by the U.S. Department of Energy, Office of Science, under contract No. DE-AC02-05CH11231.

\section*{Appendix A: Fourier Transform of nonlocal Contact Operator} \label{Appendix A}
To take the Fourier transform of the $k^2$ operator, we use a test function,
\begin{align}
    & \braket{r | \hat{O}_{C} | \psi} 
    \\
    &= \int \frac{d^3 p}{(2\pi)^3} \frac{d^3 p^{\prime}}{(2\pi)^3} d^3 r^{\prime} e^{i\bm{p}^{\prime}\cdot \bm{r}} e^{-i \bm{p}\cdot\bm{r}^{\prime}} \braket{ \bm{p}^{\prime} | \hat{O}_{C} | \bm{p}} \psi(\bm{r}^{\prime}) \nonumber
\end{align}
which, after a change of variables $\bm{p} = \bm{k} - \frac{1}{2}\bm{q}$ and ${\bm{p}^{\prime}=\bm{k} + \frac{1}{2}\bm{q}}$ gives,
\begin{align}
    \braket{r | \hat{O}_{C} | \psi} &= \int \frac{d^3 q}{(2\pi)^3} \frac{d^3 k}{(2\pi)^3} d^3 r^{\prime} \ \nonumber
    \\
    &\ \ \times \ e^{i (\bm{k} + \frac{1}{2}\bm{q}) \cdot \bm{r}} e^{-i (\bm{k} - \frac{1}{2}\bm{q})\cdot \bm{r}^{\prime}} C k^2 f(q^2) \psi(r^{\prime}) \nonumber
    \\
    &= \int \frac{d^3 q}{(2\pi)^3} \frac{d^3 k}{(2\pi)^3} d^3 r^{\prime}  C k^2 e^{i \frac{\bm{q}}{2}\cdot (\bm{r} + \bm{r}^{\prime})} \  \nonumber
    \\
    &\ \ \times e^{i\bm{k}\cdot(\bm{r} - \bm{r}^{\prime})} f(q^2) \psi(\bm{r}^{\prime}) \nonumber
    \\
    &= \int \frac{d^3 q}{(2\pi)^3} \frac{d^3 k}{(2\pi)^3} d^3 r^{\prime}  C k_{\alpha} e^{i \frac{\bm{q}}{2}\cdot (\bm{r} + \bm{r}^{\prime})}  \nonumber
    \\
    & \ \ \times i \frac{\partial}{\partial r^{\prime}_{\alpha}}e^{i\bm{k}\cdot(\bm{r} - \bm{r}^{\prime})} f(q^2) \psi(\bm{r}^{\prime})
\end{align}
where in the last line we've used $k_{\alpha}e^{i\bm{k}\cdot(\bm{r}-\bm{r}^{\prime})} = i\frac{\partial}{\partial r^{\prime}_{\alpha}}e^{i\bm{k}\cdot(\bm{r}-\bm{r}^{\prime})}$. Now we can integrate by parts over $\bm{r}^{\prime}$, taking advantage of the boundary conditions on the wavefunction to find,
\begin{align}
    \braket{r \biggl| \hat{O}_{C} \biggl| \psi} &= - \tilde{C}_3 \int \frac{d^3 q}{(2\pi)^3} \frac{d^3 k}{(2\pi)^3} d^3 r^{\prime} i k_{\alpha}  \nonumber
    \\
    &\ \ \  \times \frac{\partial}{\partial r^{\prime}_{\alpha}}\left[e^{i \frac{\bm{q}}{2}\cdot (\bm{r} + \bm{r}^{\prime})} \psi(\bm{r}^{\prime}) \right] e^{i \bm{k} \cdot (\bm{r} - \bm{r}^{\prime})} f(q^2) \nonumber
    \\
    &= - C \int \frac{d^3 q}{(2\pi)^3} \frac{d^3 k}{(2\pi)^3} d^3 r^{\prime}i \nonumber
    \\
    &\ \ \ \times \frac{\partial}{\partial r^{\prime}_{\alpha}} \left[e^{i \frac{\bm{q}}{2}\cdot (\bm{r} + \bm{r}^{\prime})} \psi(\bm{r}^{\prime}) \right] \nonumber
    \\
    &\ \ \ \times i\frac{\partial}{\partial r^{\prime}_{\alpha}}\left[e^{i \bm{k} \cdot (\bm{r} - \bm{r}^{\prime})}\right] f(q^2).
\end{align}
A second integration by parts gives,
\begin{align}
    \braket{r \biggl| \hat{O}_{\tilde{C}_3} \biggl| \psi} & = - C \int \frac{d^3 q}{(2\pi)^3} \frac{d^3 k}{(2\pi)^3} d^3 r^{\prime} \nonumber
    \\
    &\ \ \ \times \frac{\partial ^2}{\partial r^{\prime 2}_{\alpha}} \left[e^{i \frac{\bm{q}}{2}\cdot (\bm{r} + \bm{r}^{\prime})} \psi(\bm{r}^{\prime}) \right] e^{i \bm{k} \cdot (\bm{r} - \bm{r}^{\prime})} f(q^2) \nonumber
    \\
    & = - C \int \frac{d^3 q}{(2\pi)^3} \frac{d^3 k}{(2\pi)^3} d^3 r^{\prime} \nonumber
    \\
    &\ \ \ \times\frac{\partial ^2}{\partial r^{\prime 2}_{\alpha}} \left[e^{i \frac{\bm{q}}{2}\cdot \bm{r}^{\prime}} \psi(\bm{r}^{\prime}) \right] e^{i \bm{k} \cdot (\bm{r} - \bm{r}^{\prime})} f(q^2) e^{i \frac{\bm{q}}{2}\cdot \bm{r}} \nonumber
    \\
    &= - C \int \frac{d^3 q}{(2\pi)^3} \frac{d^3 k}{(2\pi)^3} d^3 r^{\prime} \biggl[ -\frac{q_{\alpha}^2}{4}e^{i \frac{\bm{q}}{2}\cdot \bm{r}^{\prime}}\psi(\bm{r}^{\prime}) \nonumber
    \\
    &\ \ \ + iq_{\alpha}e^{i \frac{\bm{q}}{2}\cdot \bm{r}^{\prime}}\frac{\partial \psi(\bm{r}^{\prime})}{\partial r_{\alpha}} + e^{i \frac{\bm{q}}{2}\cdot \bm{r}^{\prime}} \frac{\partial^2\psi(\bm{r}^{\prime})}{\partial r^{\prime 2}_{\alpha}} \biggl] \nonumber
    \\
    &\ \ \ \times e^{i \bm{k} \cdot (\bm{r} - \bm{r}^{\prime})} f(q^2) e^{i \frac{\bm{q}}{2}\cdot \bm{r}}.
\end{align}
Lastly we employ the identity $\delta(\bm{r}-\bm{r}^{\prime})= \int \frac{d^3 k}{(2\pi)^3} e^{i \bm{k}\cdot (\bm{r}-\bm{r}^{\prime})}$ and integrate over $\bm{r}^{\prime}$ to find,
\begin{align}
    \braket{r \biggl| \hat{O}_{\tilde{C}_3} \biggl| \psi} &= - C \int \frac{d^3 q}{(2\pi)^3}  \biggl[ -\frac{q_{\alpha}^2}{4}e^{i \bm{q} \cdot \bm{r}}\psi(\bm{r}) \nonumber
    \\
    & + iq_{\alpha}e^{i \bm{q} \cdot \bm{r}}\frac{\partial \psi(\bm{r})}{\partial r_{\alpha}} + e^{i \bm{q} \cdot \bm{r}} \frac{\partial^2\psi(\bm{r})}{\partial r^{ 2}_{\alpha}} \biggl]  f(q^2)  \nonumber
    \\
    &= C \biggl[ -\frac{1}{4}(\Delta \delta(\bm{r}))\psi(\bm{r}) \nonumber
    \\
    &\ \ \ - \frac{1}{r} \frac{\partial \delta(\bm{r})}{\partial r}\biggl(\bm{r} \cdot \nabla \psi(\bm{r})\biggl) - \delta(\bm{r})\Delta \psi(\bm{r}) \biggl] 
\end{align}
where for the middle term in the second line we have rewritten it using $\frac{\partial \delta(\bm{r})}{\partial r_{\alpha}} = \frac{\partial \delta(\bm{r})}{\partial r} \frac{r_{\alpha}}{r}$.
\bibliography{bibliography}

\begin{thebibliography}{61}%
\makeatletter
\providecommand \@ifxundefined [1]{%
 \@ifx{#1\undefined}
}%
\providecommand \@ifnum [1]{%
 \ifnum #1\expandafter \@firstoftwo
 \else \expandafter \@secondoftwo
 \fi
}%
\providecommand \@ifx [1]{%
 \ifx #1\expandafter \@firstoftwo
 \else \expandafter \@secondoftwo
 \fi
}%
\providecommand \natexlab [1]{#1}%
\providecommand \enquote  [1]{``#1''}%
\providecommand \bibnamefont  [1]{#1}%
\providecommand \bibfnamefont [1]{#1}%
\providecommand \citenamefont [1]{#1}%
\providecommand \href@noop [0]{\@secondoftwo}%
\providecommand \href [0]{\begingroup \@sanitize@url \@href}%
\providecommand \@href[1]{\@@startlink{#1}\@@href}%
\providecommand \@@href[1]{\endgroup#1\@@endlink}%
\providecommand \@sanitize@url [0]{\catcode `\\12\catcode `\$12\catcode
  `\&12\catcode `\#12\catcode `\^12\catcode `\_12\catcode `\%12\relax}%
\providecommand \@@startlink[1]{}%
\providecommand \@@endlink[0]{}%
\providecommand \url  [0]{\begingroup\@sanitize@url \@url }%
\providecommand \@url [1]{\endgroup\@href {#1}{\urlprefix }}%
\providecommand \urlprefix  [0]{URL }%
\providecommand \Eprint [0]{\href }%
\providecommand \doibase [0]{http://dx.doi.org/}%
\providecommand \selectlanguage [0]{\@gobble}%
\providecommand \bibinfo  [0]{\@secondoftwo}%
\providecommand \bibfield  [0]{\@secondoftwo}%
\providecommand \translation [1]{[#1]}%
\providecommand \BibitemOpen [0]{}%
\providecommand \bibitemStop [0]{}%
\providecommand \bibitemNoStop [0]{.\EOS\space}%
\providecommand \EOS [0]{\spacefactor3000\relax}%
\providecommand \BibitemShut  [1]{\csname bibitem#1\endcsname}%
\let\auto@bib@innerbib\@empty
\bibitem [{\citenamefont {Bethe}(1971)}]{Bethe_1971}%
  \BibitemOpen
  \bibfield  {author} {\bibinfo {author} {\bibfnamefont {H.~A.}\ \bibnamefont
  {Bethe}},\ }\href {\doibase 10.1146/annurev.ns.21.120171.000521} {\bibfield
  {journal} {\bibinfo  {journal} {Annu. Rev. Nucl. Part. Sci.}\ }\textbf
  {\bibinfo {volume} {21}},\ \bibinfo {pages} {93} (\bibinfo {year}
  {1971})}\BibitemShut {NoStop}%
\bibitem [{\citenamefont {Drischler}\ \emph {et~al.}(2021)\citenamefont
  {Drischler}, \citenamefont {Haxton}, \citenamefont {McElvain}, \citenamefont
  {Mereghetti}, \citenamefont {Nicholson}, \citenamefont {Vranas},\ and\
  \citenamefont {{Walker-Loud}}}]{Drischler_Haxton_McElvain_etal_2021}%
  \BibitemOpen
  \bibfield  {author} {\bibinfo {author} {\bibfnamefont {C.}~\bibnamefont
  {Drischler}}, \bibinfo {author} {\bibfnamefont {W.}~\bibnamefont {Haxton}},
  \bibinfo {author} {\bibfnamefont {K.}~\bibnamefont {McElvain}}, \bibinfo
  {author} {\bibfnamefont {E.}~\bibnamefont {Mereghetti}}, \bibinfo {author}
  {\bibfnamefont {A.}~\bibnamefont {Nicholson}}, \bibinfo {author}
  {\bibfnamefont {P.}~\bibnamefont {Vranas}}, \ and\ \bibinfo {author}
  {\bibfnamefont {A.}~\bibnamefont {{Walker-Loud}}},\ }\href {\doibase
  10.1016/j.ppnp.2021.103888} {\bibfield  {journal} {\bibinfo  {journal}
  {Progress in Particle and Nuclear Physics}\ }\textbf {\bibinfo {volume}
  {121}},\ \bibinfo {pages} {103888} (\bibinfo {year} {2021})}\BibitemShut
  {NoStop}%
\bibitem [{\citenamefont {Weinberg}(1979)}]{Weinberg_1979}%
  \BibitemOpen
  \bibfield  {author} {\bibinfo {author} {\bibfnamefont {S.}~\bibnamefont
  {Weinberg}},\ }\href {\doibase 10.1016/0378-4371(79)90223-1} {\bibfield
  {journal} {\bibinfo  {journal} {Physica A}\ }\textbf {\bibinfo {volume}
  {96}},\ \bibinfo {pages} {327} (\bibinfo {year} {1979})}\BibitemShut
  {NoStop}%
\bibitem [{\citenamefont {Weinberg}(1990)}]{Weinberg_1990}%
  \BibitemOpen
  \bibfield  {author} {\bibinfo {author} {\bibfnamefont {S.}~\bibnamefont
  {Weinberg}},\ }\href {\doibase 10.1016/0370-2693(90)90938-3} {\bibfield
  {journal} {\bibinfo  {journal} {Phys. Lett. B}\ }\textbf {\bibinfo {volume}
  {251}},\ \bibinfo {pages} {288} (\bibinfo {year} {1990})}\BibitemShut
  {NoStop}%
\bibitem [{\citenamefont {Weinberg}(1991)}]{Weinberg_1991}%
  \BibitemOpen
  \bibfield  {author} {\bibinfo {author} {\bibfnamefont {S.}~\bibnamefont
  {Weinberg}},\ }\href {\doibase 10.1016/0550-3213(91)90231-L} {\bibfield
  {journal} {\bibinfo  {journal} {Nucl. Phys. B}\ }\textbf {\bibinfo {volume}
  {363}},\ \bibinfo {pages} {3} (\bibinfo {year} {1991})}\BibitemShut {NoStop}%
\bibitem [{\citenamefont {Ord{\'o}{\~n}ez}\ \emph {et~al.}(1994)\citenamefont
  {Ord{\'o}{\~n}ez}, \citenamefont {Ray},\ and\ \citenamefont {{van
  Kolck}}}]{Ordonez_Ray_vanKolck_1994}%
  \BibitemOpen
  \bibfield  {author} {\bibinfo {author} {\bibfnamefont {C.}~\bibnamefont
  {Ord{\'o}{\~n}ez}}, \bibinfo {author} {\bibfnamefont {L.}~\bibnamefont
  {Ray}}, \ and\ \bibinfo {author} {\bibfnamefont {U.}~\bibnamefont {{van
  Kolck}}},\ }\href {\doibase 10.1103/PhysRevLett.72.1982} {\bibfield
  {journal} {\bibinfo  {journal} {Phys. Rev. Lett.}\ }\textbf {\bibinfo
  {volume} {72}},\ \bibinfo {pages} {1982} (\bibinfo {year}
  {1994})}\BibitemShut {NoStop}%
\bibitem [{\citenamefont {Ord{\'o}{\~n}ez}\ \emph {et~al.}(1996)\citenamefont
  {Ord{\'o}{\~n}ez}, \citenamefont {Ray},\ and\ \citenamefont {{van
  Kolck}}}]{Ordonez_Ray_vanKolck_1996}%
  \BibitemOpen
  \bibfield  {author} {\bibinfo {author} {\bibfnamefont {C.}~\bibnamefont
  {Ord{\'o}{\~n}ez}}, \bibinfo {author} {\bibfnamefont {L.}~\bibnamefont
  {Ray}}, \ and\ \bibinfo {author} {\bibfnamefont {U.}~\bibnamefont {{van
  Kolck}}},\ }\href {\doibase 10.1103/PhysRevC.53.2086} {\bibfield  {journal}
  {\bibinfo  {journal} {Phys. Rev. C}\ }\textbf {\bibinfo {volume} {53}},\
  \bibinfo {pages} {2086} (\bibinfo {year} {1996})}\BibitemShut {NoStop}%
\bibitem [{\citenamefont {{van Kolck}}(1994)}]{vanKolck_1994}%
  \BibitemOpen
  \bibfield  {author} {\bibinfo {author} {\bibfnamefont {U.}~\bibnamefont {{van
  Kolck}}},\ }\href {\doibase 10.1103/PhysRevC.49.2932} {\bibfield  {journal}
  {\bibinfo  {journal} {Phys. Rev. C}\ }\textbf {\bibinfo {volume} {49}},\
  \bibinfo {pages} {2932} (\bibinfo {year} {1994})}\BibitemShut {NoStop}%
\bibitem [{\citenamefont {Georgi}(1993)}]{Georgi_1993}%
  \BibitemOpen
  \bibfield  {author} {\bibinfo {author} {\bibfnamefont {H.}~\bibnamefont
  {Georgi}},\ }\href@noop {} {\bibfield  {journal} {\bibinfo  {journal} {Annu.
  Rev. Nucl. Part. Sci.}\ }\textbf {\bibinfo {volume} {43}},\ \bibinfo {pages}
  {209} (\bibinfo {year} {1993})}\BibitemShut {NoStop}%
\bibitem [{\citenamefont {Epelbaum}\ \emph {et~al.}(2009)\citenamefont
  {Epelbaum}, \citenamefont {Hammer},\ and\ \citenamefont
  {Mei{\ss}ner}}]{Epelbaum_Hammer_Meissner_2009}%
  \BibitemOpen
  \bibfield  {author} {\bibinfo {author} {\bibfnamefont {E.}~\bibnamefont
  {Epelbaum}}, \bibinfo {author} {\bibfnamefont {H.-W.}\ \bibnamefont
  {Hammer}}, \ and\ \bibinfo {author} {\bibfnamefont {U.-G.}\ \bibnamefont
  {Mei{\ss}ner}},\ }\href {\doibase 10.1103/RevModPhys.81.1773} {\bibfield
  {journal} {\bibinfo  {journal} {Rev. Mod. Phys.}\ }\textbf {\bibinfo {volume}
  {81}},\ \bibinfo {pages} {1773} (\bibinfo {year} {2009})}\BibitemShut
  {NoStop}%
\bibitem [{\citenamefont {Machleidt}\ and\ \citenamefont
  {Entem}(2011)}]{Machleidt_Entem_2011}%
  \BibitemOpen
  \bibfield  {author} {\bibinfo {author} {\bibfnamefont {R.}~\bibnamefont
  {Machleidt}}\ and\ \bibinfo {author} {\bibfnamefont {D.~R.}\ \bibnamefont
  {Entem}},\ }\href {\doibase 10.1016/j.physrep.2011.02.001} {\bibfield
  {journal} {\bibinfo  {journal} {Phys. Rep.}\ }\textbf {\bibinfo {volume}
  {503}},\ \bibinfo {pages} {1} (\bibinfo {year} {2011})}\BibitemShut {NoStop}%
\bibitem [{\citenamefont {Hagen}\ \emph {et~al.}(2008)\citenamefont {Hagen},
  \citenamefont {Papenbrock}, \citenamefont {Dean},\ and\ \citenamefont
  {{Hjorth-Jensen}}}]{Hagen_Papenbrock_Dean_etal_2008}%
  \BibitemOpen
  \bibfield  {author} {\bibinfo {author} {\bibfnamefont {G.}~\bibnamefont
  {Hagen}}, \bibinfo {author} {\bibfnamefont {T.}~\bibnamefont {Papenbrock}},
  \bibinfo {author} {\bibfnamefont {D.~J.}\ \bibnamefont {Dean}}, \ and\
  \bibinfo {author} {\bibfnamefont {M.}~\bibnamefont {{Hjorth-Jensen}}},\
  }\href {\doibase 10.1103/PhysRevLett.101.092502} {\bibfield  {journal}
  {\bibinfo  {journal} {Phys. Rev. Lett.}\ }\textbf {\bibinfo {volume} {101}},\
  \bibinfo {pages} {092502} (\bibinfo {year} {2008})}\BibitemShut {NoStop}%
\bibitem [{\citenamefont {Sun}\ \emph {et~al.}(2018)\citenamefont {Sun},
  \citenamefont {Morris}, \citenamefont {Hagen}, \citenamefont {Jansen},\ and\
  \citenamefont {Papenbrock}}]{Sun_Morris_Hagen_etal_2018}%
  \BibitemOpen
  \bibfield  {author} {\bibinfo {author} {\bibfnamefont {Z.~H.}\ \bibnamefont
  {Sun}}, \bibinfo {author} {\bibfnamefont {T.~D.}\ \bibnamefont {Morris}},
  \bibinfo {author} {\bibfnamefont {G.}~\bibnamefont {Hagen}}, \bibinfo
  {author} {\bibfnamefont {G.~R.}\ \bibnamefont {Jansen}}, \ and\ \bibinfo
  {author} {\bibfnamefont {T.}~\bibnamefont {Papenbrock}},\ }\href {\doibase
  10.1103/PhysRevC.98.054320} {\bibfield  {journal} {\bibinfo  {journal} {Phys.
  Rev. C}\ }\textbf {\bibinfo {volume} {98}},\ \bibinfo {pages} {054320}
  (\bibinfo {year} {2018})}\BibitemShut {NoStop}%
\bibitem [{\citenamefont {Barrett}\ \emph {et~al.}(2013)\citenamefont
  {Barrett}, \citenamefont {Navr{\'a}til},\ and\ \citenamefont
  {Vary}}]{Barrett_Navratil_Vary_2013}%
  \BibitemOpen
  \bibfield  {author} {\bibinfo {author} {\bibfnamefont {B.~R.}\ \bibnamefont
  {Barrett}}, \bibinfo {author} {\bibfnamefont {P.}~\bibnamefont
  {Navr{\'a}til}}, \ and\ \bibinfo {author} {\bibfnamefont {J.~P.}\
  \bibnamefont {Vary}},\ }\href {\doibase 10.1016/j.ppnp.2012.10.003}
  {\bibfield  {journal} {\bibinfo  {journal} {Prog. Nucl. Part. Phys.}\
  }\textbf {\bibinfo {volume} {69}},\ \bibinfo {pages} {131} (\bibinfo {year}
  {2013})}\BibitemShut {NoStop}%
\bibitem [{\citenamefont {Stroberg}\ \emph {et~al.}(2021)\citenamefont
  {Stroberg}, \citenamefont {Holt}, \citenamefont {Schwenk},\ and\
  \citenamefont {Simonis}}]{Stroberg_Holt_Schwenk_etal_2021}%
  \BibitemOpen
  \bibfield  {author} {\bibinfo {author} {\bibfnamefont {S.~R.}\ \bibnamefont
  {Stroberg}}, \bibinfo {author} {\bibfnamefont {J.~D.}\ \bibnamefont {Holt}},
  \bibinfo {author} {\bibfnamefont {A.}~\bibnamefont {Schwenk}}, \ and\
  \bibinfo {author} {\bibfnamefont {J.}~\bibnamefont {Simonis}},\ }\href
  {\doibase 10.1103/PhysRevLett.126.022501} {\bibfield  {journal} {\bibinfo
  {journal} {Phys. Rev. Lett.}\ }\textbf {\bibinfo {volume} {126}},\ \bibinfo
  {pages} {022501} (\bibinfo {year} {2021})}\BibitemShut {NoStop}%
\bibitem [{\citenamefont {Carlson}(1987)}]{Carlson_1987}%
  \BibitemOpen
  \bibfield  {author} {\bibinfo {author} {\bibfnamefont {J.}~\bibnamefont
  {Carlson}},\ }\href {\doibase 10.1103/PhysRevC.36.2026} {\bibfield  {journal}
  {\bibinfo  {journal} {Phys. Rev. C}\ }\textbf {\bibinfo {volume} {36}},\
  \bibinfo {pages} {2026} (\bibinfo {year} {1987})}\BibitemShut {NoStop}%
\bibitem [{\citenamefont {Schmidt}\ and\ \citenamefont
  {Fantoni}(1999)}]{Schmidt_Fantoni_1999}%
  \BibitemOpen
  \bibfield  {author} {\bibinfo {author} {\bibfnamefont {K.}~\bibnamefont
  {Schmidt}}\ and\ \bibinfo {author} {\bibfnamefont {S.}~\bibnamefont
  {Fantoni}},\ }\href {\doibase 10.1016/S0370-2693(98)01522-6} {\bibfield
  {journal} {\bibinfo  {journal} {Phys. Lett. B}\ }\textbf {\bibinfo {volume}
  {446}},\ \bibinfo {pages} {99} (\bibinfo {year} {1999})}\BibitemShut
  {NoStop}%
\bibitem [{\citenamefont {Carlson}\ \emph {et~al.}(2015)\citenamefont
  {Carlson}, \citenamefont {Gandolfi}, \citenamefont {Pederiva}, \citenamefont
  {Pieper}, \citenamefont {Schiavilla}, \citenamefont {Schmidt},\ and\
  \citenamefont {Wiringa}}]{Carlson_Gandolfi_Pederiva_etal_2015}%
  \BibitemOpen
  \bibfield  {author} {\bibinfo {author} {\bibfnamefont {J.}~\bibnamefont
  {Carlson}}, \bibinfo {author} {\bibfnamefont {S.}~\bibnamefont {Gandolfi}},
  \bibinfo {author} {\bibfnamefont {F.}~\bibnamefont {Pederiva}}, \bibinfo
  {author} {\bibfnamefont {S.~C.}\ \bibnamefont {Pieper}}, \bibinfo {author}
  {\bibfnamefont {R.}~\bibnamefont {Schiavilla}}, \bibinfo {author}
  {\bibfnamefont {K.~E.}\ \bibnamefont {Schmidt}}, \ and\ \bibinfo {author}
  {\bibfnamefont {R.~B.}\ \bibnamefont {Wiringa}},\ }\href {\doibase
  10.1103/RevModPhys.87.1067} {\bibfield  {journal} {\bibinfo  {journal} {Rev.
  Mod. Phys.}\ }\textbf {\bibinfo {volume} {87}},\ \bibinfo {pages} {1067}
  (\bibinfo {year} {2015})}\BibitemShut {NoStop}%
\bibitem [{\citenamefont {Epelbaum}\ \emph {et~al.}(2005)\citenamefont
  {Epelbaum}, \citenamefont {Gl{\"o}ckle},\ and\ \citenamefont
  {Mei{\ss}ner}}]{Epelbaum_Glockle_Meissner_2005}%
  \BibitemOpen
  \bibfield  {author} {\bibinfo {author} {\bibfnamefont {E.}~\bibnamefont
  {Epelbaum}}, \bibinfo {author} {\bibfnamefont {W.}~\bibnamefont
  {Gl{\"o}ckle}}, \ and\ \bibinfo {author} {\bibfnamefont {U.-G.}\ \bibnamefont
  {Mei{\ss}ner}},\ }\href {\doibase 10.1016/j.nuclphysa.2004.09.107} {\bibfield
   {journal} {\bibinfo  {journal} {Nucl. Phys. A}\ }\textbf {\bibinfo {volume}
  {747}},\ \bibinfo {pages} {362} (\bibinfo {year} {2005})}\BibitemShut
  {NoStop}%
\bibitem [{\citenamefont {Entem}\ and\ \citenamefont
  {Machleidt}(2003)}]{Entem_Machleidt_2003}%
  \BibitemOpen
  \bibfield  {author} {\bibinfo {author} {\bibfnamefont {D.~R.}\ \bibnamefont
  {Entem}}\ and\ \bibinfo {author} {\bibfnamefont {R.}~\bibnamefont
  {Machleidt}},\ }\href {\doibase 10.1103/PhysRevC.68.041001} {\bibfield
  {journal} {\bibinfo  {journal} {Phys. Rev. C}\ }\textbf {\bibinfo {volume}
  {68}},\ \bibinfo {pages} {041001} (\bibinfo {year} {2003})}\BibitemShut
  {NoStop}%
\bibitem [{\citenamefont {Lynn}\ and\ \citenamefont
  {Schmidt}(2012)}]{Lynn_Schmidt_2012}%
  \BibitemOpen
  \bibfield  {author} {\bibinfo {author} {\bibfnamefont {J.~E.}\ \bibnamefont
  {Lynn}}\ and\ \bibinfo {author} {\bibfnamefont {K.~E.}\ \bibnamefont
  {Schmidt}},\ }\href {\doibase 10.1103/PhysRevC.86.014324} {\bibfield
  {journal} {\bibinfo  {journal} {Phys. Rev. C}\ }\textbf {\bibinfo {volume}
  {86}},\ \bibinfo {pages} {014324} (\bibinfo {year} {2012})}\BibitemShut
  {NoStop}%
\bibitem [{\citenamefont {Roggero}\ \emph {et~al.}(2013)\citenamefont
  {Roggero}, \citenamefont {Mukherjee},\ and\ \citenamefont
  {Pederiva}}]{Roggero_Mukherjee_Pederiva_2013}%
  \BibitemOpen
  \bibfield  {author} {\bibinfo {author} {\bibfnamefont {A.}~\bibnamefont
  {Roggero}}, \bibinfo {author} {\bibfnamefont {A.}~\bibnamefont {Mukherjee}},
  \ and\ \bibinfo {author} {\bibfnamefont {F.}~\bibnamefont {Pederiva}},\
  }\href {\doibase 10.1103/PhysRevB.88.115138} {\bibfield  {journal} {\bibinfo
  {journal} {Phys. Rev. B}\ }\textbf {\bibinfo {volume} {88}},\ \bibinfo
  {pages} {115138} (\bibinfo {year} {2013})}\BibitemShut {NoStop}%
\bibitem [{\citenamefont {Gezerlis}\ \emph {et~al.}(2013)\citenamefont
  {Gezerlis}, \citenamefont {Tews}, \citenamefont {Epelbaum}, \citenamefont
  {Gandolfi}, \citenamefont {Hebeler}, \citenamefont {Nogga},\ and\
  \citenamefont {Schwenk}}]{Gezerlis_Tews_Epelbaum_etal_2013}%
  \BibitemOpen
  \bibfield  {author} {\bibinfo {author} {\bibfnamefont {A.}~\bibnamefont
  {Gezerlis}}, \bibinfo {author} {\bibfnamefont {I.}~\bibnamefont {Tews}},
  \bibinfo {author} {\bibfnamefont {E.}~\bibnamefont {Epelbaum}}, \bibinfo
  {author} {\bibfnamefont {S.}~\bibnamefont {Gandolfi}}, \bibinfo {author}
  {\bibfnamefont {K.}~\bibnamefont {Hebeler}}, \bibinfo {author} {\bibfnamefont
  {A.}~\bibnamefont {Nogga}}, \ and\ \bibinfo {author} {\bibfnamefont
  {A.}~\bibnamefont {Schwenk}},\ }\href {\doibase
  10.1103/PhysRevLett.111.032501} {\bibfield  {journal} {\bibinfo  {journal}
  {Phys. Rev. Lett.}\ }\textbf {\bibinfo {volume} {111}},\ \bibinfo {pages}
  {032501} (\bibinfo {year} {2013})}\BibitemShut {NoStop}%
\bibitem [{\citenamefont {Gezerlis}\ \emph {et~al.}(2014)\citenamefont
  {Gezerlis}, \citenamefont {Tews}, \citenamefont {Epelbaum}, \citenamefont
  {Freunek}, \citenamefont {Gandolfi}, \citenamefont {Hebeler}, \citenamefont
  {Nogga},\ and\ \citenamefont {Schwenk}}]{Gezerlis_Tews_Epelbaum_etal_2014}%
  \BibitemOpen
  \bibfield  {author} {\bibinfo {author} {\bibfnamefont {A.}~\bibnamefont
  {Gezerlis}}, \bibinfo {author} {\bibfnamefont {I.}~\bibnamefont {Tews}},
  \bibinfo {author} {\bibfnamefont {E.}~\bibnamefont {Epelbaum}}, \bibinfo
  {author} {\bibfnamefont {M.}~\bibnamefont {Freunek}}, \bibinfo {author}
  {\bibfnamefont {S.}~\bibnamefont {Gandolfi}}, \bibinfo {author}
  {\bibfnamefont {K.}~\bibnamefont {Hebeler}}, \bibinfo {author} {\bibfnamefont
  {A.}~\bibnamefont {Nogga}}, \ and\ \bibinfo {author} {\bibfnamefont
  {A.}~\bibnamefont {Schwenk}},\ }\href {\doibase 10.1103/PhysRevC.90.054323}
  {\bibfield  {journal} {\bibinfo  {journal} {Phys. Rev. C}\ }\textbf {\bibinfo
  {volume} {90}},\ \bibinfo {pages} {054323} (\bibinfo {year}
  {2014})}\BibitemShut {NoStop}%
\bibitem [{\citenamefont {Tews}\ \emph {et~al.}(2016)\citenamefont {Tews},
  \citenamefont {Gandolfi}, \citenamefont {Gezerlis},\ and\ \citenamefont
  {Schwenk}}]{Tews_Gandolfi_Gezerlis_etal_2016}%
  \BibitemOpen
  \bibfield  {author} {\bibinfo {author} {\bibfnamefont {I.}~\bibnamefont
  {Tews}}, \bibinfo {author} {\bibfnamefont {S.}~\bibnamefont {Gandolfi}},
  \bibinfo {author} {\bibfnamefont {A.}~\bibnamefont {Gezerlis}}, \ and\
  \bibinfo {author} {\bibfnamefont {A.}~\bibnamefont {Schwenk}},\ }\href
  {\doibase 10.1103/PhysRevC.93.024305} {\bibfield  {journal} {\bibinfo
  {journal} {Physical Review C}\ }\textbf {\bibinfo {volume} {93}} (\bibinfo
  {year} {2016}),\ 10.1103/PhysRevC.93.024305}\BibitemShut {NoStop}%
\bibitem [{\citenamefont {Piarulli}\ \emph {et~al.}(2015)\citenamefont
  {Piarulli}, \citenamefont {Girlanda}, \citenamefont {Schiavilla},
  \citenamefont {P{\'e}rez}, \citenamefont {Amaro},\ and\ \citenamefont
  {Arriola}}]{Piarulli_Girlanda_Schiavilla_etal_2015}%
  \BibitemOpen
  \bibfield  {author} {\bibinfo {author} {\bibfnamefont {M.}~\bibnamefont
  {Piarulli}}, \bibinfo {author} {\bibfnamefont {L.}~\bibnamefont {Girlanda}},
  \bibinfo {author} {\bibfnamefont {R.}~\bibnamefont {Schiavilla}}, \bibinfo
  {author} {\bibfnamefont {R.~N.}\ \bibnamefont {P{\'e}rez}}, \bibinfo {author}
  {\bibfnamefont {J.~E.}\ \bibnamefont {Amaro}}, \ and\ \bibinfo {author}
  {\bibfnamefont {E.~R.}\ \bibnamefont {Arriola}},\ }\href {\doibase
  10.1103/PhysRevC.91.024003} {\bibfield  {journal} {\bibinfo  {journal} {Phys.
  Rev. C}\ }\textbf {\bibinfo {volume} {91}},\ \bibinfo {pages} {024003}
  (\bibinfo {year} {2015})}\BibitemShut {NoStop}%
\bibitem [{\citenamefont {Piarulli}\ \emph {et~al.}(2016)\citenamefont
  {Piarulli}, \citenamefont {Girlanda}, \citenamefont {Schiavilla},
  \citenamefont {Kievsky}, \citenamefont {Lovato}, \citenamefont {Marcucci},
  \citenamefont {Pieper}, \citenamefont {Viviani},\ and\ \citenamefont
  {Wiringa}}]{Piarulli_Girlanda_Schiavilla_etal_2016}%
  \BibitemOpen
  \bibfield  {author} {\bibinfo {author} {\bibfnamefont {M.}~\bibnamefont
  {Piarulli}}, \bibinfo {author} {\bibfnamefont {L.}~\bibnamefont {Girlanda}},
  \bibinfo {author} {\bibfnamefont {R.}~\bibnamefont {Schiavilla}}, \bibinfo
  {author} {\bibfnamefont {A.}~\bibnamefont {Kievsky}}, \bibinfo {author}
  {\bibfnamefont {A.}~\bibnamefont {Lovato}}, \bibinfo {author} {\bibfnamefont
  {L.~E.}\ \bibnamefont {Marcucci}}, \bibinfo {author} {\bibfnamefont {S.~C.}\
  \bibnamefont {Pieper}}, \bibinfo {author} {\bibfnamefont {M.}~\bibnamefont
  {Viviani}}, \ and\ \bibinfo {author} {\bibfnamefont {R.~B.}\ \bibnamefont
  {Wiringa}},\ }\href {\doibase 10.1103/PhysRevC.94.054007} {\bibfield
  {journal} {\bibinfo  {journal} {Phys. Rev. C}\ }\textbf {\bibinfo {volume}
  {94}},\ \bibinfo {pages} {054007} (\bibinfo {year} {2016})}\BibitemShut
  {NoStop}%
\bibitem [{\citenamefont {Saha}\ \emph {et~al.}(2023)\citenamefont {Saha},
  \citenamefont {Entem}, \citenamefont {Machleidt},\ and\ \citenamefont
  {Nosyk}}]{Saha_Entem_Machleidt_etal_2023}%
  \BibitemOpen
  \bibfield  {author} {\bibinfo {author} {\bibfnamefont {S.~K.}\ \bibnamefont
  {Saha}}, \bibinfo {author} {\bibfnamefont {D.~R.}\ \bibnamefont {Entem}},
  \bibinfo {author} {\bibfnamefont {R.}~\bibnamefont {Machleidt}}, \ and\
  \bibinfo {author} {\bibfnamefont {Y.}~\bibnamefont {Nosyk}},\ }\href
  {\doibase 10.1103/PhysRevC.107.034002} {\bibfield  {journal} {\bibinfo
  {journal} {Phys. Rev. C}\ }\textbf {\bibinfo {volume} {107}},\ \bibinfo
  {pages} {034002} (\bibinfo {year} {2023})}\BibitemShut {NoStop}%
\bibitem [{\citenamefont {Somasundaram}\ \emph {et~al.}(2024)\citenamefont
  {Somasundaram}, \citenamefont {Lynn}, \citenamefont {Huth}, \citenamefont
  {Schwenk},\ and\ \citenamefont {Tews}}]{Somasundaram_Lynn_Huth_etal_2024}%
  \BibitemOpen
  \bibfield  {author} {\bibinfo {author} {\bibfnamefont {R.}~\bibnamefont
  {Somasundaram}}, \bibinfo {author} {\bibfnamefont {J.~E.}\ \bibnamefont
  {Lynn}}, \bibinfo {author} {\bibfnamefont {L.}~\bibnamefont {Huth}}, \bibinfo
  {author} {\bibfnamefont {A.}~\bibnamefont {Schwenk}}, \ and\ \bibinfo
  {author} {\bibfnamefont {I.}~\bibnamefont {Tews}},\ }\href {\doibase
  10.1103/PhysRevC.109.034005} {\bibfield  {journal} {\bibinfo  {journal}
  {Phys. Rev. C}\ }\textbf {\bibinfo {volume} {109}},\ \bibinfo {pages}
  {034005} (\bibinfo {year} {2024})}\BibitemShut {NoStop}%
\bibitem [{\citenamefont {Curry}\ \emph {et~al.}(2023)\citenamefont {Curry},
  \citenamefont {Lynn}, \citenamefont {Schmidt},\ and\ \citenamefont
  {Gezerlis}}]{Curry_Lynn_Schmidt_etal_2023}%
  \BibitemOpen
  \bibfield  {author} {\bibinfo {author} {\bibfnamefont {R.}~\bibnamefont
  {Curry}}, \bibinfo {author} {\bibfnamefont {J.~E.}\ \bibnamefont {Lynn}},
  \bibinfo {author} {\bibfnamefont {K.~E.}\ \bibnamefont {Schmidt}}, \ and\
  \bibinfo {author} {\bibfnamefont {A.}~\bibnamefont {Gezerlis}},\ }\href
  {\doibase 10.1103/PhysRevResearch.5.L042021} {\bibfield  {journal} {\bibinfo
  {journal} {Phys. Rev. Res.}\ }\textbf {\bibinfo {volume} {5}},\ \bibinfo
  {pages} {L042021} (\bibinfo {year} {2023})}\BibitemShut {NoStop}%
\bibitem [{\citenamefont {Bogner}\ \emph {et~al.}(2010)\citenamefont {Bogner},
  \citenamefont {Furnstahl},\ and\ \citenamefont
  {Schwenk}}]{Bogner_Furnstahl_Schwenk_2010}%
  \BibitemOpen
  \bibfield  {author} {\bibinfo {author} {\bibfnamefont {S.~K.}\ \bibnamefont
  {Bogner}}, \bibinfo {author} {\bibfnamefont {R.~J.}\ \bibnamefont
  {Furnstahl}}, \ and\ \bibinfo {author} {\bibfnamefont {A.}~\bibnamefont
  {Schwenk}},\ }\href {\doibase 10.1016/j.ppnp.2010.03.001} {\bibfield
  {journal} {\bibinfo  {journal} {Prog. Part. Nucl. Phys.}\ }\textbf {\bibinfo
  {volume} {65}},\ \bibinfo {pages} {94} (\bibinfo {year} {2010})}\BibitemShut
  {NoStop}%
\bibitem [{\citenamefont {Tews}\ \emph {et~al.}(2024)\citenamefont {Tews},
  \citenamefont {Somasundaram}, \citenamefont {Lonardoni}, \citenamefont
  {G{\"o}ttling}, \citenamefont {Seutin}, \citenamefont {Carlson},
  \citenamefont {Gandolfi}, \citenamefont {Hebeler},\ and\ \citenamefont
  {Schwenk}}]{Tews_Somasundaram_Lonardoni_etal_2024}%
  \BibitemOpen
  \bibfield  {author} {\bibinfo {author} {\bibfnamefont {I.}~\bibnamefont
  {Tews}}, \bibinfo {author} {\bibfnamefont {R.}~\bibnamefont {Somasundaram}},
  \bibinfo {author} {\bibfnamefont {D.}~\bibnamefont {Lonardoni}}, \bibinfo
  {author} {\bibfnamefont {H.}~\bibnamefont {G{\"o}ttling}}, \bibinfo {author}
  {\bibfnamefont {R.}~\bibnamefont {Seutin}}, \bibinfo {author} {\bibfnamefont
  {J.}~\bibnamefont {Carlson}}, \bibinfo {author} {\bibfnamefont
  {S.}~\bibnamefont {Gandolfi}}, \bibinfo {author} {\bibfnamefont
  {K.}~\bibnamefont {Hebeler}}, \ and\ \bibinfo {author} {\bibfnamefont
  {A.}~\bibnamefont {Schwenk}},\ }\href {\doibase 10.48550/arXiv.2407.08979} {\
   (\bibinfo {year} {2024}),\ 10.48550/arXiv.2407.08979}\BibitemShut {NoStop}%
\bibitem [{\citenamefont {Stoks}\ \emph {et~al.}(1993)\citenamefont {Stoks},
  \citenamefont {Klomp}, \citenamefont {Rentmeester},\ and\ \citenamefont
  {De~Swart}}]{Stoks_Klomp_Rentmeester_etal_1993}%
  \BibitemOpen
  \bibfield  {author} {\bibinfo {author} {\bibfnamefont {V.~G.~J.}\
  \bibnamefont {Stoks}}, \bibinfo {author} {\bibfnamefont {R.~A.~M.}\
  \bibnamefont {Klomp}}, \bibinfo {author} {\bibfnamefont {M.~C.~M.}\
  \bibnamefont {Rentmeester}}, \ and\ \bibinfo {author} {\bibfnamefont {J.~J.}\
  \bibnamefont {De~Swart}},\ }\href {\doibase 10.1103/PhysRevC.48.792}
  {\bibfield  {journal} {\bibinfo  {journal} {Phys. Rev. C}\ }\textbf {\bibinfo
  {volume} {48}},\ \bibinfo {pages} {792} (\bibinfo {year} {1993})}\BibitemShut
  {NoStop}%
\bibitem [{\citenamefont {Epelbaum}\ \emph {et~al.}(2004)\citenamefont
  {Epelbaum}, \citenamefont {Gl{\"o}ckle},\ and\ \citenamefont
  {Mei{\ss}ner}}]{Epelbaum_Glockle_Meissner_2004}%
  \BibitemOpen
  \bibfield  {author} {\bibinfo {author} {\bibfnamefont {E.}~\bibnamefont
  {Epelbaum}}, \bibinfo {author} {\bibfnamefont {W.}~\bibnamefont
  {Gl{\"o}ckle}}, \ and\ \bibinfo {author} {\bibfnamefont {U.-G.}\ \bibnamefont
  {Mei{\ss}ner}},\ }\href {\doibase 10.1140/epja/i2003-10096-0} {\bibfield
  {journal} {\bibinfo  {journal} {Eur. Phys. J. A}\ }\textbf {\bibinfo {volume}
  {19}},\ \bibinfo {pages} {125} (\bibinfo {year} {2004})}\BibitemShut
  {NoStop}%
\bibitem [{\citenamefont {Huth}(2018)}]{Huth_2018}%
  \BibitemOpen
  \bibfield  {author} {\bibinfo {author} {\bibfnamefont {L.}~\bibnamefont
  {Huth}},\ }\emph {\bibinfo {title} {Local Interactions and Shell-Model
  Interactions from Chiral Effective Field Theory}},\ \href@noop {} {Ph.D.
  thesis} (\bibinfo {year} {2018})\BibitemShut {NoStop}%
\bibitem [{\citenamefont {Ceperley}(1996)}]{Ceperley_1996a}%
  \BibitemOpen
  \bibfield  {author} {\bibinfo {author} {\bibfnamefont {D.~M.}\ \bibnamefont
  {Ceperley}},\ }\href {\doibase 10.1103/RevModPhys.67.279} {\bibfield
  {journal} {\bibinfo  {journal} {Rev. Mod. Phys.}\ }\textbf {\bibinfo {volume}
  {67}},\ \bibinfo {pages} {279} (\bibinfo {year} {1996})}\BibitemShut
  {NoStop}%
\bibitem [{\citenamefont {Koloren{\v c}}\ and\ \citenamefont
  {Mitas}(2011)}]{Kolorenc_Mitas_2011}%
  \BibitemOpen
  \bibfield  {author} {\bibinfo {author} {\bibfnamefont {J.}~\bibnamefont
  {Koloren{\v c}}}\ and\ \bibinfo {author} {\bibfnamefont {L.}~\bibnamefont
  {Mitas}},\ }\href {\doibase 10.1088/0034-4885/74/2/026502} {\bibfield
  {journal} {\bibinfo  {journal} {Rep. Prog. Phys.}\ }\textbf {\bibinfo
  {volume} {74}},\ \bibinfo {pages} {026502} (\bibinfo {year}
  {2011})}\BibitemShut {NoStop}%
\bibitem [{\citenamefont {Giorgini}\ \emph {et~al.}(2008)\citenamefont
  {Giorgini}, \citenamefont {Pitaevskii},\ and\ \citenamefont
  {Stringari}}]{Giorgini_Pitaevskii_Stringari_2008}%
  \BibitemOpen
  \bibfield  {author} {\bibinfo {author} {\bibfnamefont {S.}~\bibnamefont
  {Giorgini}}, \bibinfo {author} {\bibfnamefont {L.~P.}\ \bibnamefont
  {Pitaevskii}}, \ and\ \bibinfo {author} {\bibfnamefont {S.}~\bibnamefont
  {Stringari}},\ }\href {\doibase 10.1103/RevModPhys.80.1215} {\bibfield
  {journal} {\bibinfo  {journal} {Rev. Mod. Phys.}\ }\textbf {\bibinfo {volume}
  {80}},\ \bibinfo {pages} {1215} (\bibinfo {year} {2008})}\BibitemShut
  {NoStop}%
\bibitem [{\citenamefont {Gandolfi}\ \emph {et~al.}(2011)\citenamefont
  {Gandolfi}, \citenamefont {Schmidt},\ and\ \citenamefont
  {Carlson}}]{Gandolfi_Schmidt_Carlson_2011}%
  \BibitemOpen
  \bibfield  {author} {\bibinfo {author} {\bibfnamefont {S.}~\bibnamefont
  {Gandolfi}}, \bibinfo {author} {\bibfnamefont {K.~E.}\ \bibnamefont
  {Schmidt}}, \ and\ \bibinfo {author} {\bibfnamefont {J.}~\bibnamefont
  {Carlson}},\ }\href {\doibase 10.1103/PhysRevA.83.041601} {\bibfield
  {journal} {\bibinfo  {journal} {Phys. Rev. A}\ }\textbf {\bibinfo {volume}
  {83}},\ \bibinfo {pages} {041601} (\bibinfo {year} {2011})}\BibitemShut
  {NoStop}%
\bibitem [{\citenamefont {Gezerlis}\ and\ \citenamefont
  {Carlson}(2010)}]{Gezerlis_Carlson_2010}%
  \BibitemOpen
  \bibfield  {author} {\bibinfo {author} {\bibfnamefont {A.}~\bibnamefont
  {Gezerlis}}\ and\ \bibinfo {author} {\bibfnamefont {J.}~\bibnamefont
  {Carlson}},\ }\href {\doibase 10.1103/PhysRevC.81.025803} {\bibfield
  {journal} {\bibinfo  {journal} {Phys. Rev. C}\ }\textbf {\bibinfo {volume}
  {81}},\ \bibinfo {pages} {025803} (\bibinfo {year} {2010})}\BibitemShut
  {NoStop}%
\bibitem [{\citenamefont {Lynn}\ \emph {et~al.}(2016)\citenamefont {Lynn},
  \citenamefont {Tews}, \citenamefont {Carlson}, \citenamefont {Gandolfi},
  \citenamefont {Gezerlis}, \citenamefont {Schmidt},\ and\ \citenamefont
  {Schwenk}}]{Lynn_Tews_Carlson_etal_2016}%
  \BibitemOpen
  \bibfield  {author} {\bibinfo {author} {\bibfnamefont {J.~E.}\ \bibnamefont
  {Lynn}}, \bibinfo {author} {\bibfnamefont {I.}~\bibnamefont {Tews}}, \bibinfo
  {author} {\bibfnamefont {J.}~\bibnamefont {Carlson}}, \bibinfo {author}
  {\bibfnamefont {S.}~\bibnamefont {Gandolfi}}, \bibinfo {author}
  {\bibfnamefont {A.}~\bibnamefont {Gezerlis}}, \bibinfo {author}
  {\bibfnamefont {K.~E.}\ \bibnamefont {Schmidt}}, \ and\ \bibinfo {author}
  {\bibfnamefont {A.}~\bibnamefont {Schwenk}},\ }\href {\doibase
  10.1103/PhysRevLett.116.062501} {\bibfield  {journal} {\bibinfo  {journal}
  {Phys. Rev. Lett.}\ }\textbf {\bibinfo {volume} {116}},\ \bibinfo {pages}
  {062501} (\bibinfo {year} {2016})}\BibitemShut {NoStop}%
\bibitem [{\citenamefont {Lynn}\ \emph {et~al.}(2017)\citenamefont {Lynn},
  \citenamefont {Tews}, \citenamefont {Carlson}, \citenamefont {Gandolfi},
  \citenamefont {Gezerlis}, \citenamefont {Schmidt},\ and\ \citenamefont
  {Schwenk}}]{Lynn_Tews_Carlson_etal_2017}%
  \BibitemOpen
  \bibfield  {author} {\bibinfo {author} {\bibfnamefont {J.~E.}\ \bibnamefont
  {Lynn}}, \bibinfo {author} {\bibfnamefont {I.}~\bibnamefont {Tews}}, \bibinfo
  {author} {\bibfnamefont {J.}~\bibnamefont {Carlson}}, \bibinfo {author}
  {\bibfnamefont {S.}~\bibnamefont {Gandolfi}}, \bibinfo {author}
  {\bibfnamefont {A.}~\bibnamefont {Gezerlis}}, \bibinfo {author}
  {\bibfnamefont {K.~E.}\ \bibnamefont {Schmidt}}, \ and\ \bibinfo {author}
  {\bibfnamefont {A.}~\bibnamefont {Schwenk}},\ }\href {\doibase
  10.1103/PhysRevC.96.054007} {\bibfield  {journal} {\bibinfo  {journal} {Phys.
  Rev. C}\ }\textbf {\bibinfo {volume} {96}},\ \bibinfo {pages} {054007}
  (\bibinfo {year} {2017})}\BibitemShut {NoStop}%
\bibitem [{\citenamefont {Lynn}\ \emph {et~al.}(2019)\citenamefont {Lynn},
  \citenamefont {Tews}, \citenamefont {Gandolfi},\ and\ \citenamefont
  {Lovato}}]{Lynn_Tews_Gandolfi_etal_2019}%
  \BibitemOpen
  \bibfield  {author} {\bibinfo {author} {\bibfnamefont {J.}~\bibnamefont
  {Lynn}}, \bibinfo {author} {\bibfnamefont {I.}~\bibnamefont {Tews}}, \bibinfo
  {author} {\bibfnamefont {S.}~\bibnamefont {Gandolfi}}, \ and\ \bibinfo
  {author} {\bibfnamefont {A.}~\bibnamefont {Lovato}},\ }\href {\doibase
  10.1146/annurev-nucl-101918-023600} {\bibfield  {journal} {\bibinfo
  {journal} {Annu. Rev. Nucl. Part. Sci.}\ }\textbf {\bibinfo {volume} {69}},\
  \bibinfo {pages} {279} (\bibinfo {year} {2019})}\BibitemShut {NoStop}%
\bibitem [{\citenamefont {Curry}\ \emph {et~al.}(2024)\citenamefont {Curry},
  \citenamefont {Dissanayake}, \citenamefont {Gandolfi},\ and\ \citenamefont
  {Gezerlis}}]{Curry_Dissanayake_Gandolfi_etal_2024}%
  \BibitemOpen
  \bibfield  {author} {\bibinfo {author} {\bibfnamefont {R.}~\bibnamefont
  {Curry}}, \bibinfo {author} {\bibfnamefont {J.}~\bibnamefont {Dissanayake}},
  \bibinfo {author} {\bibfnamefont {S.}~\bibnamefont {Gandolfi}}, \ and\
  \bibinfo {author} {\bibfnamefont {A.}~\bibnamefont {Gezerlis}},\ }\href
  {\doibase 10.1098/rsta.2023.0127} {\bibfield  {journal} {\bibinfo  {journal}
  {Phil. Trans. R. Soc. A.}\ }\textbf {\bibinfo {volume} {382}},\ \bibinfo
  {pages} {20230127} (\bibinfo {year} {2024})}\BibitemShut {NoStop}%
\bibitem [{\citenamefont {Lynn}\ \emph {et~al.}(2014)\citenamefont {Lynn},
  \citenamefont {Carlson}, \citenamefont {Epelbaum}, \citenamefont {Gandolfi},
  \citenamefont {Gezerlis},\ and\ \citenamefont
  {Schwenk}}]{Lynn_Carlson_Epelbaum_etal_2014}%
  \BibitemOpen
  \bibfield  {author} {\bibinfo {author} {\bibfnamefont {J.~E.}\ \bibnamefont
  {Lynn}}, \bibinfo {author} {\bibfnamefont {J.}~\bibnamefont {Carlson}},
  \bibinfo {author} {\bibfnamefont {E.}~\bibnamefont {Epelbaum}}, \bibinfo
  {author} {\bibfnamefont {S.}~\bibnamefont {Gandolfi}}, \bibinfo {author}
  {\bibfnamefont {A.}~\bibnamefont {Gezerlis}}, \ and\ \bibinfo {author}
  {\bibfnamefont {A.}~\bibnamefont {Schwenk}},\ }\href {\doibase
  10.1103/PhysRevLett.113.192501} {\bibfield  {journal} {\bibinfo  {journal}
  {Phys. Rev. Lett.}\ }\textbf {\bibinfo {volume} {113}},\ \bibinfo {pages}
  {192501} (\bibinfo {year} {2014})}\BibitemShut {NoStop}%
\bibitem [{\citenamefont {Lonardoni}\ \emph
  {et~al.}(2018{\natexlab{a}})\citenamefont {Lonardoni}, \citenamefont
  {Carlson}, \citenamefont {Gandolfi}, \citenamefont {Lynn}, \citenamefont
  {Schmidt}, \citenamefont {Schwenk},\ and\ \citenamefont
  {Wang}}]{Lonardoni_Carlson_Gandolfi_etal_2018}%
  \BibitemOpen
  \bibfield  {author} {\bibinfo {author} {\bibfnamefont {D.}~\bibnamefont
  {Lonardoni}}, \bibinfo {author} {\bibfnamefont {J.}~\bibnamefont {Carlson}},
  \bibinfo {author} {\bibfnamefont {S.}~\bibnamefont {Gandolfi}}, \bibinfo
  {author} {\bibfnamefont {J.~E.}\ \bibnamefont {Lynn}}, \bibinfo {author}
  {\bibfnamefont {K.~E.}\ \bibnamefont {Schmidt}}, \bibinfo {author}
  {\bibfnamefont {A.}~\bibnamefont {Schwenk}}, \ and\ \bibinfo {author}
  {\bibfnamefont {X.~B.}\ \bibnamefont {Wang}},\ }\href {\doibase
  10.1103/PhysRevLett.120.122502} {\bibfield  {journal} {\bibinfo  {journal}
  {Phys. Rev. Lett.}\ }\textbf {\bibinfo {volume} {120}},\ \bibinfo {pages}
  {122502} (\bibinfo {year} {2018}{\natexlab{a}})}\BibitemShut {NoStop}%
\bibitem [{\citenamefont {Sarsa}\ \emph {et~al.}(2003)\citenamefont {Sarsa},
  \citenamefont {Fantoni}, \citenamefont {Schmidt},\ and\ \citenamefont
  {Pederiva}}]{Sarsa_Fantoni_Schmidt_etal_2003}%
  \BibitemOpen
  \bibfield  {author} {\bibinfo {author} {\bibfnamefont {A.}~\bibnamefont
  {Sarsa}}, \bibinfo {author} {\bibfnamefont {S.}~\bibnamefont {Fantoni}},
  \bibinfo {author} {\bibfnamefont {K.~E.}\ \bibnamefont {Schmidt}}, \ and\
  \bibinfo {author} {\bibfnamefont {F.}~\bibnamefont {Pederiva}},\ }\href
  {\doibase 10.1103/PhysRevC.68.024308} {\bibfield  {journal} {\bibinfo
  {journal} {Phys. Rev. C}\ }\textbf {\bibinfo {volume} {68}},\ \bibinfo
  {pages} {024308} (\bibinfo {year} {2003})}\BibitemShut {NoStop}%
\bibitem [{\citenamefont {Gezerlis}\ and\ \citenamefont
  {Carlson}(2008)}]{Gezerlis_Carlson_2008}%
  \BibitemOpen
  \bibfield  {author} {\bibinfo {author} {\bibfnamefont {A.}~\bibnamefont
  {Gezerlis}}\ and\ \bibinfo {author} {\bibfnamefont {J.}~\bibnamefont
  {Carlson}},\ }\href {\doibase 10.1103/PhysRevC.77.032801} {\bibfield
  {journal} {\bibinfo  {journal} {Phys. Rev. C}\ }\textbf {\bibinfo {volume}
  {77}},\ \bibinfo {pages} {032801} (\bibinfo {year} {2008})}\BibitemShut
  {NoStop}%
\bibitem [{\citenamefont {Gandolfi}\ \emph {et~al.}(2015)\citenamefont
  {Gandolfi}, \citenamefont {Gezerlis},\ and\ \citenamefont
  {Carlson}}]{Gandolfi_Gezerlis_Carlson_2015}%
  \BibitemOpen
  \bibfield  {author} {\bibinfo {author} {\bibfnamefont {S.}~\bibnamefont
  {Gandolfi}}, \bibinfo {author} {\bibfnamefont {A.}~\bibnamefont {Gezerlis}},
  \ and\ \bibinfo {author} {\bibfnamefont {J.}~\bibnamefont {Carlson}},\ }\href
  {\doibase 10.1146/annurev-nucl-102014-021957} {\bibfield  {journal} {\bibinfo
   {journal} {Annu. Rev. Nucl. Part. Sci.}\ }\textbf {\bibinfo {volume} {65}},\
  \bibinfo {pages} {303} (\bibinfo {year} {2015})}\BibitemShut {NoStop}%
\bibitem [{\citenamefont {Gandolfi}\ \emph {et~al.}(2022)\citenamefont
  {Gandolfi}, \citenamefont {Palkanoglou}, \citenamefont {Carlson},
  \citenamefont {Gezerlis},\ and\ \citenamefont
  {Schmidt}}]{Gandolfi_Palkanoglou_Carlson_etal_2022}%
  \BibitemOpen
  \bibfield  {author} {\bibinfo {author} {\bibfnamefont {S.}~\bibnamefont
  {Gandolfi}}, \bibinfo {author} {\bibfnamefont {G.}~\bibnamefont
  {Palkanoglou}}, \bibinfo {author} {\bibfnamefont {J.}~\bibnamefont
  {Carlson}}, \bibinfo {author} {\bibfnamefont {A.}~\bibnamefont {Gezerlis}}, \
  and\ \bibinfo {author} {\bibfnamefont {K.~E.}\ \bibnamefont {Schmidt}},\
  }\href {\doibase 10.3390/condmat7010019} {\bibfield  {journal} {\bibinfo
  {journal} {Condes. Matter}\ }\textbf {\bibinfo {volume} {7}},\ \bibinfo
  {pages} {19} (\bibinfo {year} {2022})}\BibitemShut {NoStop}%
\bibitem [{\citenamefont {Gandolfi}\ \emph {et~al.}(2014)\citenamefont
  {Gandolfi}, \citenamefont {Lovato}, \citenamefont {Carlson},\ and\
  \citenamefont {Schmidt}}]{Gandolfi_Lovato_Carlson_etal_2014}%
  \BibitemOpen
  \bibfield  {author} {\bibinfo {author} {\bibfnamefont {S.}~\bibnamefont
  {Gandolfi}}, \bibinfo {author} {\bibfnamefont {A.}~\bibnamefont {Lovato}},
  \bibinfo {author} {\bibfnamefont {J.}~\bibnamefont {Carlson}}, \ and\
  \bibinfo {author} {\bibfnamefont {K.~E.}\ \bibnamefont {Schmidt}},\ }\href
  {\doibase 10.1103/PhysRevC.90.061306} {\bibfield  {journal} {\bibinfo
  {journal} {Phys. Rev. C}\ }\textbf {\bibinfo {volume} {90}},\ \bibinfo
  {pages} {061306} (\bibinfo {year} {2014})}\BibitemShut {NoStop}%
\bibitem [{\citenamefont {Lonardoni}\ \emph {et~al.}(2020)\citenamefont
  {Lonardoni}, \citenamefont {Tews}, \citenamefont {Gandolfi},\ and\
  \citenamefont {Carlson}}]{Lonardoni_Tews_Gandolfi_etal_2020}%
  \BibitemOpen
  \bibfield  {author} {\bibinfo {author} {\bibfnamefont {D.}~\bibnamefont
  {Lonardoni}}, \bibinfo {author} {\bibfnamefont {I.}~\bibnamefont {Tews}},
  \bibinfo {author} {\bibfnamefont {S.}~\bibnamefont {Gandolfi}}, \ and\
  \bibinfo {author} {\bibfnamefont {J.}~\bibnamefont {Carlson}},\ }\href
  {\doibase 10.1103/PhysRevResearch.2.022033} {\bibfield  {journal} {\bibinfo
  {journal} {Phys. Rev. Res.}\ }\textbf {\bibinfo {volume} {2}},\ \bibinfo
  {pages} {022033} (\bibinfo {year} {2020})}\BibitemShut {NoStop}%
\bibitem [{\citenamefont {Pudliner}\ \emph {et~al.}(1997)\citenamefont
  {Pudliner}, \citenamefont {Pandharipande}, \citenamefont {Carlson},
  \citenamefont {Pieper},\ and\ \citenamefont
  {Wiringa}}]{Pudliner_Pandharipande_Carlson_etal_1997}%
  \BibitemOpen
  \bibfield  {author} {\bibinfo {author} {\bibfnamefont {B.~S.}\ \bibnamefont
  {Pudliner}}, \bibinfo {author} {\bibfnamefont {V.~R.}\ \bibnamefont
  {Pandharipande}}, \bibinfo {author} {\bibfnamefont {J.}~\bibnamefont
  {Carlson}}, \bibinfo {author} {\bibfnamefont {S.~C.}\ \bibnamefont {Pieper}},
  \ and\ \bibinfo {author} {\bibfnamefont {R.~B.}\ \bibnamefont {Wiringa}},\
  }\href {\doibase 10.1103/PhysRevC.56.1720} {\bibfield  {journal} {\bibinfo
  {journal} {Phys. Rev. C}\ }\textbf {\bibinfo {volume} {56}},\ \bibinfo
  {pages} {1720} (\bibinfo {year} {1997})}\BibitemShut {NoStop}%
\bibitem [{\citenamefont {King}\ and\ \citenamefont
  {Pastore}(2024)}]{King_Pastore_2024}%
  \BibitemOpen
  \bibfield  {author} {\bibinfo {author} {\bibfnamefont {G.~B.}\ \bibnamefont
  {King}}\ and\ \bibinfo {author} {\bibfnamefont {S.}~\bibnamefont {Pastore}},\
  }\href@noop {} {\bibfield  {journal} {\bibinfo  {journal} {Annu. Rev. Nucl.
  Part. Sci.}\ }\textbf {\bibinfo {volume} {74}} (\bibinfo {year} {2024})},\
  \Eprint {http://arxiv.org/abs/2402.06602} {arXiv:2402.06602 [nucl-th]}
  \BibitemShut {NoStop}%
\bibitem [{\citenamefont {Lonardoni}\ \emph
  {et~al.}(2018{\natexlab{b}})\citenamefont {Lonardoni}, \citenamefont
  {Gandolfi}, \citenamefont {Lynn}, \citenamefont {Petrie}, \citenamefont
  {Carlson}, \citenamefont {Schmidt},\ and\ \citenamefont
  {Schwenk}}]{Lonardoni_Gandolfi_Lynn_etal_2018}%
  \BibitemOpen
  \bibfield  {author} {\bibinfo {author} {\bibfnamefont {D.}~\bibnamefont
  {Lonardoni}}, \bibinfo {author} {\bibfnamefont {S.}~\bibnamefont {Gandolfi}},
  \bibinfo {author} {\bibfnamefont {J.~E.}\ \bibnamefont {Lynn}}, \bibinfo
  {author} {\bibfnamefont {C.}~\bibnamefont {Petrie}}, \bibinfo {author}
  {\bibfnamefont {J.}~\bibnamefont {Carlson}}, \bibinfo {author} {\bibfnamefont
  {K.~E.}\ \bibnamefont {Schmidt}}, \ and\ \bibinfo {author} {\bibfnamefont
  {A.}~\bibnamefont {Schwenk}},\ }\href {\doibase 10.1103/PhysRevC.97.044318}
  {\bibfield  {journal} {\bibinfo  {journal} {Phys. Rev. C}\ }\textbf {\bibinfo
  {volume} {97}},\ \bibinfo {pages} {044318} (\bibinfo {year}
  {2018}{\natexlab{b}})}\BibitemShut {NoStop}%
\bibitem [{\citenamefont {Pandharipande}\ and\ \citenamefont
  {Wiringa}(1979)}]{Pandharipande_Wiringa_1979}%
  \BibitemOpen
  \bibfield  {author} {\bibinfo {author} {\bibfnamefont {V.~R.}\ \bibnamefont
  {Pandharipande}}\ and\ \bibinfo {author} {\bibfnamefont {R.~B.}\ \bibnamefont
  {Wiringa}},\ }\href {\doibase 10.1103/RevModPhys.51.821} {\bibfield
  {journal} {\bibinfo  {journal} {Rev. Mod. Phys.}\ }\textbf {\bibinfo {volume}
  {51}},\ \bibinfo {pages} {821} (\bibinfo {year} {1979})}\BibitemShut
  {NoStop}%
\bibitem [{\citenamefont {Sorella}(2001)}]{Sorella_2001}%
  \BibitemOpen
  \bibfield  {author} {\bibinfo {author} {\bibfnamefont {S.}~\bibnamefont
  {Sorella}},\ }\href {\doibase 10.1103/PhysRevB.64.024512} {\bibfield
  {journal} {\bibinfo  {journal} {Phys. Rev. B}\ }\textbf {\bibinfo {volume}
  {64}},\ \bibinfo {pages} {024512} (\bibinfo {year} {2001})}\BibitemShut
  {NoStop}%
\bibitem [{\citenamefont {Lu}\ \emph {et~al.}(2022)\citenamefont {Lu},
  \citenamefont {Li}, \citenamefont {Elhatisari}, \citenamefont {Ma},
  \citenamefont {Lee},\ and\ \citenamefont
  {Mei{\ss}ner}}]{Lu_Li_Elhatisari_etal_2022}%
  \BibitemOpen
  \bibfield  {author} {\bibinfo {author} {\bibfnamefont {B.-N.}\ \bibnamefont
  {Lu}}, \bibinfo {author} {\bibfnamefont {N.}~\bibnamefont {Li}}, \bibinfo
  {author} {\bibfnamefont {S.}~\bibnamefont {Elhatisari}}, \bibinfo {author}
  {\bibfnamefont {Y.-Z.}\ \bibnamefont {Ma}}, \bibinfo {author} {\bibfnamefont
  {D.}~\bibnamefont {Lee}}, \ and\ \bibinfo {author} {\bibfnamefont {U.-G.}\
  \bibnamefont {Mei{\ss}ner}},\ }\href {\doibase
  10.1103/PhysRevLett.128.242501} {\bibfield  {journal} {\bibinfo  {journal}
  {Phys. Rev. Lett.}\ }\textbf {\bibinfo {volume} {128}},\ \bibinfo {pages}
  {242501} (\bibinfo {year} {2022})}\BibitemShut {NoStop}%
\bibitem [{\citenamefont {Foulkes}\ \emph {et~al.}(2001)\citenamefont
  {Foulkes}, \citenamefont {Mitas}, \citenamefont {Needs},\ and\ \citenamefont
  {Rajagopal}}]{Foulkes_Mitas_Needs_etal_2001}%
  \BibitemOpen
  \bibfield  {author} {\bibinfo {author} {\bibfnamefont {W.~M.~C.}\
  \bibnamefont {Foulkes}}, \bibinfo {author} {\bibfnamefont {L.}~\bibnamefont
  {Mitas}}, \bibinfo {author} {\bibfnamefont {R.~J.}\ \bibnamefont {Needs}}, \
  and\ \bibinfo {author} {\bibfnamefont {G.}~\bibnamefont {Rajagopal}},\ }\href
  {\doibase 10.1103/RevModPhys.73.33} {\bibfield  {journal} {\bibinfo
  {journal} {Rev. Mod. Phys.}\ }\textbf {\bibinfo {volume} {73}},\ \bibinfo
  {pages} {33} (\bibinfo {year} {2001})}\BibitemShut {NoStop}%
\bibitem [{\citenamefont {Drischler}\ \emph {et~al.}(2020)\citenamefont
  {Drischler}, \citenamefont {Furnstahl}, \citenamefont {Melendez},\ and\
  \citenamefont {Phillips}}]{Drischler_Furnstahl_Melendez_etal_2020}%
  \BibitemOpen
  \bibfield  {author} {\bibinfo {author} {\bibfnamefont {C.}~\bibnamefont
  {Drischler}}, \bibinfo {author} {\bibfnamefont {R.~J.}\ \bibnamefont
  {Furnstahl}}, \bibinfo {author} {\bibfnamefont {J.~A.}\ \bibnamefont
  {Melendez}}, \ and\ \bibinfo {author} {\bibfnamefont {D.~R.}\ \bibnamefont
  {Phillips}},\ }\href {\doibase 10.1103/PhysRevLett.125.202702} {\bibfield
  {journal} {\bibinfo  {journal} {Phys. Rev. Lett.}\ }\textbf {\bibinfo
  {volume} {125}},\ \bibinfo {pages} {202702} (\bibinfo {year}
  {2020})}\BibitemShut {NoStop}%
\bibitem [{\citenamefont {Gandolfi}\ \emph {et~al.}(2009)\citenamefont
  {Gandolfi}, \citenamefont {Illarionov}, \citenamefont {Schmidt},
  \citenamefont {Pederiva},\ and\ \citenamefont
  {Fantoni}}]{Gandolfi_Illarionov_Schmidt_etal_2009}%
  \BibitemOpen
  \bibfield  {author} {\bibinfo {author} {\bibfnamefont {S.}~\bibnamefont
  {Gandolfi}}, \bibinfo {author} {\bibfnamefont {A.~{\relax Yu}.}\ \bibnamefont
  {Illarionov}}, \bibinfo {author} {\bibfnamefont {K.~E.}\ \bibnamefont
  {Schmidt}}, \bibinfo {author} {\bibfnamefont {F.}~\bibnamefont {Pederiva}}, \
  and\ \bibinfo {author} {\bibfnamefont {S.}~\bibnamefont {Fantoni}},\ }\href
  {\doibase 10.1103/PhysRevC.79.054005} {\bibfield  {journal} {\bibinfo
  {journal} {Phys. Rev. C}\ }\textbf {\bibinfo {volume} {79}},\ \bibinfo
  {pages} {054005} (\bibinfo {year} {2009})}\BibitemShut {NoStop}%
\end{thebibliography}%
\end{document}